%% file: CompFrac.tex
\documentclass{article}
\pdfoutput=1 
\usepackage{arxiv}

\usepackage{framed,multirow}
\usepackage[utf8]{inputenc} 
\usepackage[T1]{fontenc}    
\usepackage{hyperref}       
\usepackage{url}            
\usepackage{booktabs}       
\usepackage{amsfonts}       
\usepackage{nicefrac}       
\usepackage{microtype}      
\usepackage{lipsum}
\usepackage{xcolor}
\usepackage{tikz}
\usepackage{amsmath}
\usepackage{amssymb}
\usepackage{siunitx}
\usepackage{lineno}
\usepackage{booktabs}
\usepackage[square,numbers]{natbib}
\usepackage{hypernat}

\newcommand{\todo}[1]{{\bf \color{blue} #1}}

\newcommand{\aaline}[1]{{AA$^\prime$}}
\newcommand{\bbline}[1]{{BB$^\prime$}}

\title{Comparison and Application of non-Conforming Mesh Models for Flow in Fractured Porous Media using dual {L}agrange multipliers}

\author{
{Patrick~Zulian} \\
Institute~of~Computational~Science,~USI~Lugano,~6904~Lugano,~Switzerland \\
\texttt{patrick.zulian@usi.ch}
\AND
{Philipp~Sch\"{a}dle}\\
Institute of Geophysics, Department of Earth Sciences,~ETH~Z\"urich,~8092~Z\"urich,~Switzerland. \\
\texttt{philipp.schaedle@erdw.ethz.ch}
\AND
{Liudmila Karagyaur}\\
Institute~of~Computational~Science,~USI~Lugano,~6904~Lugano,~Switzerland \\
\texttt{liudmila.karagyaur@usi.ch}
\AND
{Maria~G.~C.~Nestola}\\
Institute~of~Computational~Science,~USI~Lugano,~6904~Lugano,~Switzerland. \\
Institute of Geochemistry and Petrology, Department of Earth Sciences,~ETH~Z\"urich,~8092~Z\"urich,~Switzerland \\
\texttt{nestom@usi.ch}, \texttt{maria.nestola@erdw.ethz.ch}
}

\begin{document}
\maketitle
\begin{abstract}
Geological settings with reservoir characteristics include fractures with different material and geometrical properties. Hence,  numerical simulations in applied geophysics demands for computational frameworks which efficiently allow to integrate various fracture geometries in a porous medium matrix.
This study presents a modeling approach for single-phase flow in fractured porous media and its application to different types of non-conforming mesh models. 
We propose a combination of the Lagrange multiplier method with variational transfer to allow for complex non-conforming geometries as well as hybrid- and equi-dimensional models and discretizations of flow through fractured porous media.
The variational transfer is based on the $L^2$-projection and enables an accurate and highly efficient parallel projection of fields between non-conforming meshes (e.g.,\ between fracture and porous matrix domain). \\
We present the different techniques as a unified mathematical framework with a practical perspective.
By means of numerical examples we discuss both, performance and applicability of the particular strategies.
Comparisons of finite element simulation results to widely adopted 2D benchmark cases show good agreement and the dual Lagrange multiplier spaces show good performance.
In an extension to 3D fracture networks, we first provide complementary results to a recently developed benchmark case, before we explore a complex scenario which leverages the different types of fracture meshes.
Complex and highly conductive fracture networks are found more suitable in combination with embedded hybrid-dimensional fractures. 
However, thick and blocking fractures are better approximated by equi-dimensional embedded fractures and the equi-dimensional mortar method, respectively. 
\end{abstract}

%
%


\section{Introduction} 
\label{sec:intro}
%
Fluid flow through fractured porous media is a crucial process in the context of numerous subsurface applications, e.g.\ groundwater management, geothermal energy utilization, CO$_2$ sequestration, hazardous waste storage, and enhanced oil and gas recovery~\cite{pochon_2008,read_2013,tester2006,mcclure2014b,bond2003,bonnet2001,rasmuson1986,amann2018}.
Often, fluid flow velocities in fractures and in the porous medium matrix range over many orders of magnitudes.
Therefore, single fractures and networks of fractures largely govern the characteristics of fluid transport in fracture-dominated porous media.
More specifically, the detailed fracture geometry of each individual fracture has a significant influence on the fluid flow in a fracture-dominated system.
While the aperture width of a fracture can vary over several orders of magnitudes, the fracture can additionally be filled with an infilling material exhibiting permeability values ranging over many orders of magnitude.
As a consequence, these two aspects lead to strong differences in the transmissivity and therefore the fractures ability to permit fluid transport.
Thereby, the presence and permeability of an infilling material determines if a single fracture acts as a conduit or a barrier for fluid flow~\cite{dreuzy2012,zimmerman1991,ebigbo2016,vogler2016}.
A detailed description of fluid flow through fractured porous media therefore requires comprehensive knowledge about the hydraulic properties of each individual fracture. 
Such properties are very difficult to obtain in the field and a detailed deterministic description of such systems is not possible~\cite{neuman_1997}.
Consequently, stochastic investigations are universally conducted to describe fractured media and to account for associated uncertainties~\cite{berkowitz2002,dreuzy2012}.
Due to the large number of forward simulations required by stochastic studies, highly efficient and accurate numerical methods and mesh generation approaches are needed~\cite{cacas1990,hobe2018,neuman2005,dreuzy2013,dessirier2018}.\\
To numerically model fractured systems, two method classes are widely used, continuum models and discrete fracture models.
In the class of continuum models, fractures and porous-medium matrix are represented by separate continua within the same geometric mesh~\cite{warren1963,barenblatt1960,kazemi1969,kazemi1976}. 
Effective flow properties are obtained by upscaling and information between the continua needs to be transferred. 
In the class of discrete fracture models, fractures are represented as discrete domains in a numerical mesh~\cite{noorishad1982,baca1984}.
Here, two concepts are distinguished where the porous-medium matrix is either represented in discrete-fracture-matrix models (DFM) or neglected in discrete-fracture-networks models (DFN)~\cite{hyman2015,dreuzy2012}.
In classic DFMs, fracture and porous-medium matrix domains are explicitly meshed and conforming at the domain interface, i.e.\ conforming geometry and discretization~\cite{flemisch_2011,lipnikov_2014,lee_2015,lee_2019}. 
Due to the large complexity of fracture networks, mesh generation for such conforming DFMs can be very challenging and time consuming~\cite{cacace2015,holm2006,blessent2009}.
Such challenges occur in the matrix mesh domain as well as the fracture mesh domain. 
Due to the large length-to-width ratio of many fractures it is common to use lower-dimensional elements to represent the fracture domain, e.g.~\cite{karimi2003,bogdanov2003,monteagudo2004,helmig1997}. 
This avoids elements with large aspect ratios in the fracture mesh and thus improves numerical performance.
However, fractures with considerably large aperture width are not ideally represented by lower-dimensional elements~\cite{flemisch2018}.
Those fractures might be less numerous but require to generate fracture domain meshes which are equi-dimensional to the porous medium matrix domain.
Further challenges are posed by the matrix mesh generation around the fractures.
This is particularly difficult where fractures are close to each other or intersect with very low angles. 
Such configurations can lead to elements with large aspect ratios or non-physical connections, which requires fine tuning of the meshes to improve performance and stability of the solution~\cite{flemisch2018}.
It is important to bear in mind that this is even more challenging in 3D and makes stochastic studies with DFMs very challenging. 
These challenges in model generation combined with the requirements and geometrical complexities of DFMs motivated a large number of method developments and improvements,  and yielding this to be an active research field.\\
The drawbacks associated with the classic discrete-domain approaches have triggered research on, and the development of, numerical methods that allow to use individual meshes for the fracture and porous medium matrix domains.
Such methods rely on the concept of non-conforming discretizations, while the meshes might be geometrically conforming or non-conforming. 
Methods with non-conforming discretization but conforming geometries require the element facets of the fracture domain to align with the neighboring element facets of the porous-medium matrix domain without coinciding, e.g.\, mortar methods~\cite{keilegavlen_2019,nordbotten_2019,devloo_2019,duran_2019,brenner_2016b,brenner_2016} and discontinuos Galerking methods~\cite{antonietti2019mixed}.
In contrast, fully non-conforming methods, i.e.\, non-conforming discretization and non-conforming geometry, require no geometrical relationship between the fracture and the porous-medium matrix domains.
These methods exist for finite volume schemes, e.g.\ (p)EDFM \cite{li2008,hajibeygi2011,tene2017,moinfar2014,nikitin_2020} and for finite element schemes, e.g.\ extended finite element methods (XFEM)~\cite{flemisch2016}, or continuous Galerkin method where fractures are represented as Dirac functions~\cite{xu_2020}.
XFEM approaches exist with primal formulations~\cite{capatina2016,huang2011,schwenck2015} and dual formulations~\cite{dangelo2012,fumagalli2013}. An exposition of different coupling techniques in discrete fracture networks is provided in~\cite{fumagalli2019conforming}, where the focus is on handling complex fracture networks without accounting the effects of the surrounding rock.
Recently,~\citet{koeppel2018a} presented an alternative fully non-conforming finite element formulation for which~\citet{SCHADLE201942} demonstrated its applicability in 3D.
Often, non-conforming mesh methods allow for an automated mesh generation and model setup process, which enables stochastic studies with a large number of different fracture network geometries. 
In particular, methods that handle fracture and matrix meshes separately (i.e.\ non-conforming geometries), significantly reduce the work required for preparing the simulation geometries.
\citet{berre2018} provide a good overview of existing conceptual and discretization methods and discuss the differences in conforming and non-conforming methods.\\
The mutually non-conforming discretizations of non-conforming mesh approaches require the use of coupling techniques and information transfer between the domains. 
Lagrange multipliers are a common tool to couple systems of equations and they are widely used throughout various fields~\cite{NESTOLA2019108884,osborn2018}.
More specifically, fictitious domain~\cite{glowinski1994} and mortar~\cite{bernardi1993domain} methods are discretization techniques based on the Lagrange multiplier~\cite{popp2012,vonPlanta2018}. 
In the context of DFMs, Lagrange multipliers have thus far been applied for both, fully non-conforming and non-conforming discretizations with conforming geometries.
\citet{frih2012} and~\citet{boon2018} presented approaches where the porous-matrix domain is split along the fracture planes. 
The resulting sub-domains are then meshed independently and glued 
together using the mortar method. 
Alternatively, discontinuous Galerkin methods are used to handle non-conforming interfaces between any pair of elements, hence providing flexibility at a finer granularity with respect to the mortar method~\cite{antonietti2019mixed}. 
A fully non-conforming approach using Lagrange multipliers to couple fracture and porous matrix discretizations is studied by~\citet{koeppel2018a} and~\citet{SCHADLE201942}.
Particularly in~\cite{SCHADLE201942}, the $L^2$-projection with different discrete Lagrange multipliers are used to transfer information between the fracture and porous-medium matrix domains in a variationally consistent way.
Both,~\citet{koeppel2018a} and~\citet{SCHADLE201942} studied flow through DFMs with fractures of codimension one. 
While these hybrid-dimensional DFMs are widely applied and very efficient with respect to meshing and numerical performance they are less suited for fractures with large aperture widths~\cite{flemisch2018}.
For such cases, equi-dimensional descriptions of the porous-medium matrix and the fractures provide more accurate results. 
However, this requires a volume-to-volume coupling between the fracture and the matrix discretizations.
Consequently, the $L^2$-projection needs to be constructed considering volumetric polyhedral intersections.
Volumetric coupling combined with variational transfer has been studied for several applications related to fluid-structure interaction~\cite{HESCH2014853,NESTOLA2019108884}. 
Note that, as for the hybrid-dimensional case, local mass conservation is generally not guaranteed and jumping pressure coefficients can not be represented properly with our continuous Galerkin approach and Lagrange finite element spaces. 
One underlying assumptions of the Lagrange multiplier approach is a continuity of the solution across the fractures, which prevents them to act as a flow barriers.
However, some fractures are populated with an infilling material with very low permeability and therefore hinder fluid flow. 
To model such fractures, equi-dimensional sub-domains with very low permeability values need to be employed and then coupled to the porous-medium matrix using the mortar method. 
Furthermore, \citet{SCHADLE201942} found that the solution is less accurate in areas with fracture intersections and boundaries, which is more enhanced by steep pressure gradients located in these areas. Local adaptive mesh refinement would improve accuracy in these cases.
\citet{SCHADLE201942} employed dual-Lagrange multipliers~\cite{popp2012,wohlmuth2000}, which have shown to have a positive impact on the condition number and further allow to construct a symmetric positive-definite system of equations. Solving such systems is much more convenient than solving saddle-point problems, which are indefinite systems and are typically harder to solve. 
These types of systems can be solved with a wide range of methods (e.g., conjugate gradient method) and preconditioners (e.g., multigrid~\cite{briggs2000}), and enable to perform large scale computations in a convenient manner.  
Furthermore, the dual-Lagrange mulitplier space facilitates combining different coupling strategies, i.e.\, fully non-conforming hybrid-dimensional and equi-dimensional as well as mortar coupling, in a simple and unified way. \\
This paper presents a unified framework based on the method of Lagrange multipliers, which combines embedded discretization methods with non-conforming domain decomposition approaches. Embedded discretization methods are designed to couple overlapping meshes which are mutually non-conforming and can have non-matching geometric features. Here, the finite element discretization of the matrix is coupled with any number of fracture discretizations, which can be hybrid-dimensional and equi-dimensional.
Non-conforming domain decomposition techniques, such as mortar, allow us to split the domain into multiple sub-domains, discretize them independently, and couple them at their interfaces.
\\
This combination allows to employ each technique to scenarios where it is most suited, i.e.\, blocking fractures, fractures with large apertures, or fractures with large aspect ratio in networks with many fractures.
By choosing the dual Lagrange multipliers space for each of the aforementioned coupling approaches, the arising systems of equations are easily combined and condensed, as mentioned earlier, into a unique symmetric positive-definite matrix.
We show how non-conforming adaptive mesh refinement is combined with the variational transfer, and we employ it to control the error as well as to reduce the number of degrees of freedom in the arising system of equations.
We study the approaches both in isolation and combined.  In particular, the accuracy and performance of the presented frameworks is demonstrated by comparison to standard benchmark cases in 2D and 3D as well as a realistic scenario which combines the different approaches.
\\
In Section~\ref{sec:method} we describe the overall methodology. We illustrate the unified formulation of the flow model (Section~\ref{sec:formulation}), its variational formulation (Section~\ref{sec:varform}), the finite element discretization (Section~\ref{sec:disc}), the necessary steps for coupling the different discretizations with dual Lagrange multipliers (Section~\ref{sec:transfer}), and the construction of the algebraic system of equations.
In Section~\ref{sec:refine}, we show how non-conforming mesh refinement can be integrated within the coupling framework, followed by some specific details about the implementation in Section~\ref{sec:implementation}.
Numerical investigations and experiments are illustrated and discussed in Section~\ref{sec:results}. 
Finally, a conclusion of our findings and future developments are provided in Section~\ref{sec:conclusion}.

\section{Method}\label{sec:method}
\begin{figure}[ht]\centering\footnotesize
    \includegraphics[width=0.5\linewidth]{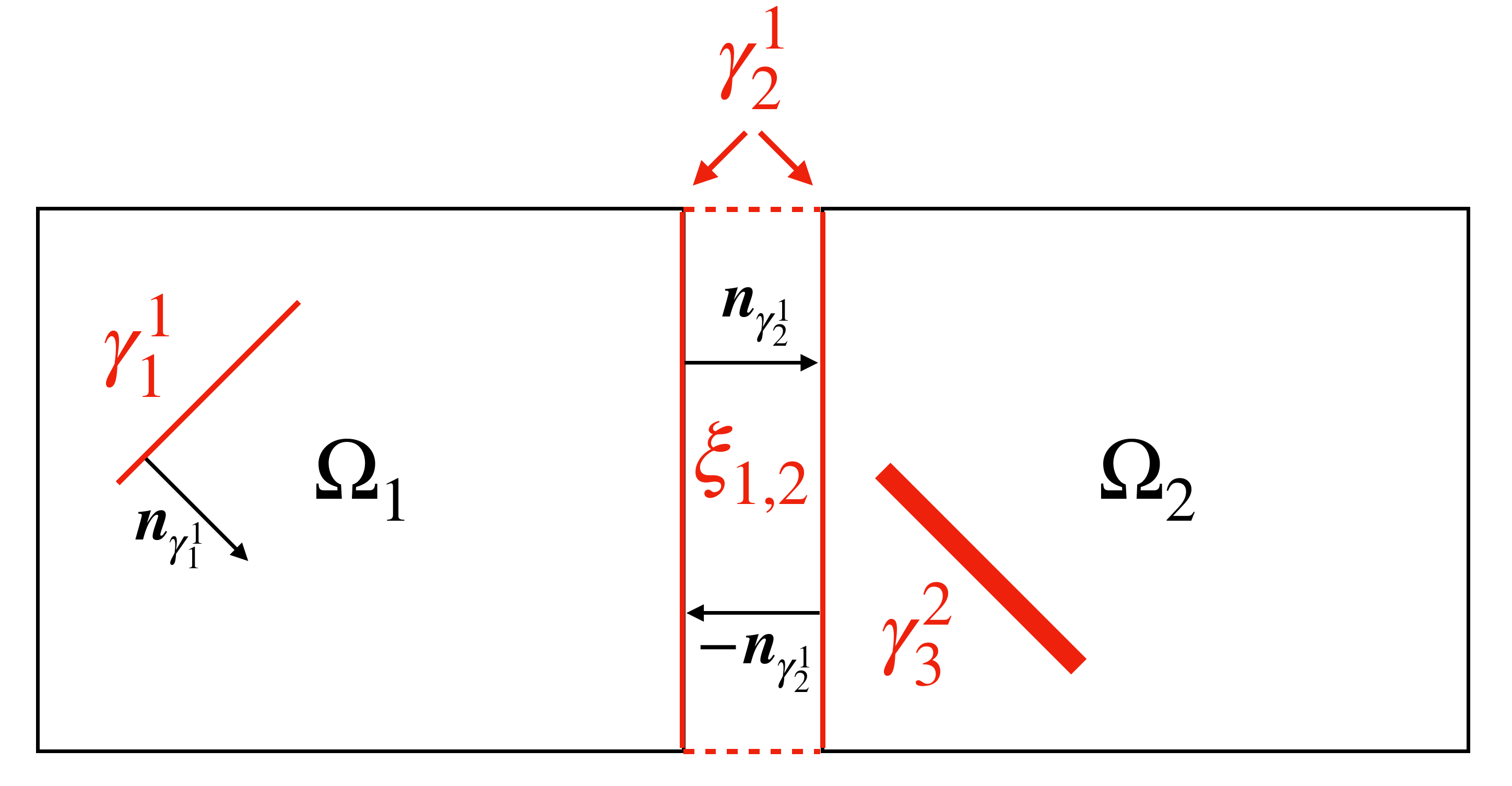}
    \caption{Two-dimensional example with an embedded lower-dimensional fracture $\gamma^1_1$, a lower dimensional (line) fracture $\gamma^1_2$ at the interface $\xi_{1, 2}$, between sub-domains $\Omega_1$ and $\Omega_2$ of the matrix, and an embedded equi-dimensional (polygon) fracture $\gamma^2_3$. The dashed lines represent an nonexistent spacing which is employed only for illustrative purposes.}
    \label{fig:geom}
\end{figure}

The method of Lagrange multipliers allows to discretize flow problems for porous media with two main types of non-conformity. 
First, the matrix is split into sub-domains which can be discretized independently then glued together using the mortar method~\cite{wohlmuth2000}. Not only the sub-domains can differ in terms of permeability, but their interface can represent fractures. 
Second, the fractures are represented as separate bodies embedded in the matrix. Such fractures are described by either lower-dimensional manifolds or equi-dimensional manifolds. 
In this section, we present a unified framework to describe the different geometric representations depicted in Figure~\ref{fig:geom}, and the related discretization techniques.

%
\subsection{Problem formulation}\label{sec:formulation}
Let $\Omega \subset \mathbb{R}^d, d \in \{2, 3\}$ be the matrix domain
with the following decomposition into $N$ sub-domains
$$
\Omega = \bigcup_{i=1}^N \Omega_i,
$$
where $\Omega_i \cap \Omega_j = \emptyset, i \neq j$. If $\Omega_i$ and $\Omega_j$ are connected, hence $\overline{\Omega}_i \cap \overline{\Omega}_j \neq \emptyset$, their interface is described by
$\xi_{ij} = \partial \Omega_i \cap \partial\Omega_j \cap \Omega$. With $\Xi = \{ \xi_{ij} \}$.

Let $\gamma \subset \Omega$ be a manifold of dimension $d$ or $d-1$ describing the fracture domains with the following decomposition
$$
\gamma = 
\bigcup_{k=1}^{N_\gamma} \gamma_k.
$$
When required we distinguish the dimension of the manifold $\gamma_k$, with
$\gamma_k^{d-1}$ we have a lower-dimensional fracture, and with $\gamma_k^{d}$ an equi-dimensional fracture.
If the interface $\xi = \xi_{ij}$ is interpreted as a lower-dimensional fracture we employ the short-hand notation $\gamma_\xi = \gamma^{d-1}_\xi$. 

Steady state fluid flow in the matrix $\Omega$ is governed by
\begin{equation}  \label{eq:mat}
    \nabla \cdot (-\boldsymbol{K}\nabla p) - \lambda = f \qquad \mathrm{in}\quad \Omega,
\end{equation}
with $p = \overline{p}$ on $\partial\Omega_D$,
where  $p$ is the pressure, $\boldsymbol{K} \in \mathbb{R}^{d \times d}$ is a bounded symmetric positive definite permeability tensor,  $f$ is a sink/source term, $\overline{p}$ is a given pressure on the boundary $\partial\Omega$ of the domain of interest $\Omega$. 

Flow in the fracture-network $\gamma$ is described by
\begin{equation} \label{eq:frac}
    \nabla \cdot (-\boldsymbol{K}_\gamma\nabla p_\gamma) + \lambda = f _\gamma \qquad  \mathrm{in} \quad  \gamma,
\end{equation}
with $p_\gamma = \overline{p}_\gamma$ on $\partial\gamma_D$.

The fluid exchange between $\Omega$ and $\gamma$ is given by the Lagrange multiplier $\lambda \in \Lambda(\gamma)$.
The function spaces $V_\Omega$ and $V_\gamma$  are defined by
\begin{align*}
V_\Omega = H^1(\Omega), & \quad W_{\Omega} = H^1_0(\Omega), \\
V_\gamma = H^1(\gamma), & \quad W_{\gamma} = H^1_0(\gamma), \\
V = V_\Omega \times V_\gamma, & \quad W = W_\Omega \times W_\gamma,
\end{align*}
where $H^1$ is the Sobolev space of weakly differentiable functions, and $H^1_0 \subset H^1$ its restriction to functions vanishing at the boundary. The Lagrange multiplier space is defined as the dual of $W$ with the product
$$
\Lambda = \prod_{k=1}^{N_{\gamma^{d-1}}} H^{-\frac{1}{2}}_{00}(\gamma^{d-1}_k) \times \prod_{k=1}^{N_{\gamma^{d}}} H^{-1}_{00}(\gamma^{d}_k), \qquad N_\gamma = N_{\gamma^{d-1}} + N_{\gamma^{d}}.
$$


Note that the Lagrange multiplier is extended by zero outside $\gamma$.
For each $\gamma_k \in \gamma$ the corresponding Lagrange multiplier is denoted with $\lambda_k \in \Lambda_k = \Lambda(\gamma_k)$.

Depending on the type of fracture the pressure term and the Lagrange multiplier have slightly different meanings.
For the lower-dimensional fracture, i.e., $\gamma^{d-1}_k$ the pressure term $p_{\gamma_k}$ represents the average pressure across the fracture with tangential permeability $\boldsymbol{K}_{\gamma_k}$, and the Lagrange multiplier 
$$\lambda_{k} = [\boldsymbol{K}\nabla p \cdot \boldsymbol{n}_{\gamma_k}] = 
(\boldsymbol{K}\nabla p - \boldsymbol{K}_{\gamma_k}\nabla p_{\gamma_k}) \cdot \boldsymbol{n}_{\gamma_k}
\quad p = p(\boldsymbol{x}), \boldsymbol{x} \in \gamma_k^{d-1}$$
represents the jump of the fluid pressure gradient in normal direction $\boldsymbol{n}_{\gamma_k}$ with respect to the fracture surface $\gamma_k$.

For the embedded equi-dimensional fracture the Lagrange multiplier can be thought as a reactive force field introduced in order to ensure the continuity of the pressure.

For a more compact notation, the aperture of lower dimensional fractures is neglected and considered in the permeability tensor.

\subsection{Weak formulation}\label{sec:varform}
 With $(\cdot, \cdot)_\Omega$ and  $(\cdot, \cdot)_\gamma$ we denote the $L^2$-inner product over $\Omega$ and $\gamma$, respectively. 
 The variational formulation is found by multiplying~\eqref{eq:mat} and \eqref{eq:frac} by test functions and integrating over the domains $\Omega$ and $\gamma$, using integration by parts. Hence, the weak form of the coupled system of equations is given as follows: find $(p, p_\gamma) \in V$ and $\lambda \in \Lambda$, such that
 \begin{equation} \label{eq:varform}
     (\boldsymbol{K} \nabla p, \nabla q)_\Omega + (\boldsymbol{K}_\gamma \nabla p_\gamma, \nabla q_\gamma)_\gamma
     - (\lambda, q - q_\gamma)_\gamma = (f, q)_\Omega + (f_\gamma, q_\gamma)_\gamma
     \quad \forall (q, q_\gamma) \in W,
 \end{equation}
 and the weak equality condition
 \begin{equation} \label{eq:weakeq}
     (p - p_\gamma, \mu)_\gamma \qquad \forall \mu \in \Lambda,
 \end{equation}
 are satisfied.
 
\subsection{Discretization} \label{sec:disc}
The variational formulation introduced in Section~\ref{sec:varform} is discretized using the finite element method. 
Depending on the settings different meshes $\mathcal{M}_i = \mathcal{M}_{\Omega_i}$ and $\mathcal{M}_{\gamma_k}$ are used for the different sub-domains of the porous matrix $\Omega_i, i = 1, \hdots N$, and for the fractures $\gamma_k, k = 1 \hdots N_\gamma$ respectively. 
The presented techniques and their implementation allow for an arbitrary choice of $\mathcal{M}_{\lambda_k}$, which is the mesh associated with the Lagrange multiplier, however since we restrict ourselves to a particular choice of multiplier space, we set $\mathcal{M}_{\lambda_k} = \mathcal{M}_{\gamma_k}$.

The method allows for a wide variety of elements for each of the meshes. 
For a manifold with dimension $d$ we employ either Lagrange elements $\mathbb{P}^k$, or tensor-product elements $\mathbb{Q}^{k}$ of order $k \in \{1, 2\}$
\begin{equation}\label{eq:fespace}
\begin{aligned}
W_{h,\alpha} = \{ & w \in W(\alpha) \colon \forall E \in \mathcal{M}_\alpha, \\
& w |_{E} \in 
    \left \{ 
        \begin{aligned} 
            \mathbb{P}^k   &\quad \text{if}~E~\text{is a simplex} \\
            \mathbb{Q}^{k} &\quad \text{if}~E~\text{is a hyper-cuboid}
        \end{aligned} 
    \right \}
    \\
\}, \\
& \alpha \in \{ \Omega, \gamma^d, \gamma^{d-1} \} \\
\Lambda_{h,\beta}  \phantom{ = \{ } & \beta \in \{ \gamma^d, \gamma^{d-1} \},
\end{aligned}
\end{equation}
where $\Lambda_h$ is a discrete Lagrange multiplier space. 
Let $\{ \varphi_i \}_{i \in J}$ be a basis of $W_{h,\Omega}$, 
$\{ \theta_j \}_{j \in J_\gamma}$ a basis of $W_{h,\gamma}$, 
and $\{ \psi_k \}_{k \in J_\gamma}$ a basis of  $\Lambda_h$, where $J$ and $J_\gamma \subset \mathbb{N}$ are index sets of the node-sets of their respective meshes $\mathcal{M}$ and $\mathcal{M}_\gamma$.
Writing the functions $q \in W_{h,\Omega}$,  $q_\gamma \in W_{h,\gamma}$, and $\mu \in \Lambda_{h,\gamma}$ in terms of their respective bases and coefficients, they read $q = \sum_{i \in J} q_i \varphi_i$, $q_\gamma = \sum_{j \in J_\gamma} q_{j\gamma} \theta_j$, and $\mu = \sum_{k \in J_\gamma} \mu_{k\gamma} \psi_k$.

The dual shape functions $\psi_j \in \Lambda_h(\gamma)$ are constructed in such a way that they satisfy the bi-orthogonality condition~\cite{wohlmuth2000}:
\begin{equation} \label{eq:dualmult}
     (\theta_i, \psi_j)_{\gamma_h} = \delta_{ij}  (\theta_i, 1)_{\gamma_h} \qquad \forall i,j \in J_{\gamma},
\end{equation} 
and integral positivity
\begin{equation} \label{eq:intpositive}
    (\psi_j, 1)_{\gamma_h} > 0.
\end{equation}
Note that~\eqref{eq:intpositive} is naturally satisfied for first order finite elements. For second order elements we follow the construction described in~\cite{popp2012,lamichhane2005}.

After reformulating the variational problem~\eqref{eq:varform} as a set of point-wise equations, the discrete problem for the porous matrix reads:
\begin{equation}\label{eq:matsys}
\sum_{i \in J} p_i (\mathbf{K} \nabla \varphi_i, \nabla \varphi_j)_{\Omega_h} 
- \sum_{k \in J_\gamma} \lambda_k  (\psi_k, \varphi_j)_{\gamma_h}
= (f,  \varphi_j)_{\Omega_h}, \qquad \forall j \in J,
\end{equation}
which translates to the linear system $\mathbf{A} \mathbf{p} - \mathbf{B}^T \boldsymbol{\lambda} = \mathbf{f}$.The fracture equations result in,
 \begin{equation}\label{eq:fracsys}
 \sum_{i \in J} p_{i,\gamma} (\mathbf{K}_\gamma \nabla \theta_i, \nabla \theta_j)_{\gamma_h}
 + \sum_{k \in J_\gamma} \lambda_k (\psi_k, \theta_j)_{\gamma_h}
 =  (f_\gamma, \theta_j)_{\gamma_h}, \qquad \forall j \in J_\gamma,
 \end{equation}
which translates to the linear system $\mathbf{A}_\gamma \mathbf{p}_\gamma + \mathbf{D}^T \boldsymbol{\lambda} = \mathbf{f}_\gamma$. 
 The weak-equality condition~\eqref{eq:weakeq} results in,
 \begin{equation} \label{eq:lagrsys}
    - \left ( \sum_{i \in J} p_i (\varphi_i, \psi_j)_{\Omega_h}  -  \sum_{k \in J} p_{k,\gamma} (\mathbf\theta_k, \psi_j)_{\gamma_h} \right) = 0,  \qquad \forall j \in J_\lambda,
 \end{equation}
which translates to the linear system $-\mathbf{B} \mathbf{p} + \mathbf{D} \mathbf{p}_\gamma = \mathbf{0}$. 
The discretization of the complete problem results in saddle-point system as follows:
 \begin{equation} 
 \label{eq:saddlepoint}
 \left|
 \begin{array}{ccc}
     \mathbf{A} & \mathbf{0}        & -\mathbf{B}^T \\
     \mathbf{0} & \mathbf{A}_\gamma & \mathbf{D}^T \\
     -\mathbf{B} & \mathbf{D}       & \mathbf{0}
 \end{array}
 \right|
 \left|
 \begin{array}{l}
 \mathbf{p} \\
 \mathbf{p}_\gamma \\
 \boldsymbol{\lambda}
 \end{array}
 \right|
 =
 \left|
 \begin{array}{l}
 \mathbf{f} \\
 \mathbf{f}_\gamma \\
 \mathbf{0}
 \end{array}
 \right|,
 \end{equation}
 However, the trivially invertible matrix $\mathbf{D}$ enables us to perform block Gaussian elimination and obtain the following statically condensed system~\cite{paz2001}
\begin{equation} 
\label{eq:condensed}
    (\mathbf{A}  +  \mathbf{T}^T \mathbf{A}_\gamma \mathbf{T}) \mathbf{p} = \mathbf{f} + \mathbf{T}^T \mathbf{f}_\gamma,
\end{equation}
where $\mathbf{T} = \mathbf{D}^{-1}\mathbf{B}$. 
Once the system is solved for $\mathbf{p}$, the solution for the fracture network can be computed by $\mathbf{p}_\gamma = \mathbf{T}\mathbf{p}$. The resulting system matrix is symmetric positive definite which allows us to adopt optimal solution strategies such as Multigrid methods~\cite{briggs2000}.

\subsection{Handling multiple types of non-conforming mesh interactions} \label{sec:multimesh}
\begin{figure}[ht]\centering\footnotesize
    \parbox{0.8\linewidth}{
        \parbox{0.48\linewidth}{\centering
            \includegraphics[width=\linewidth]{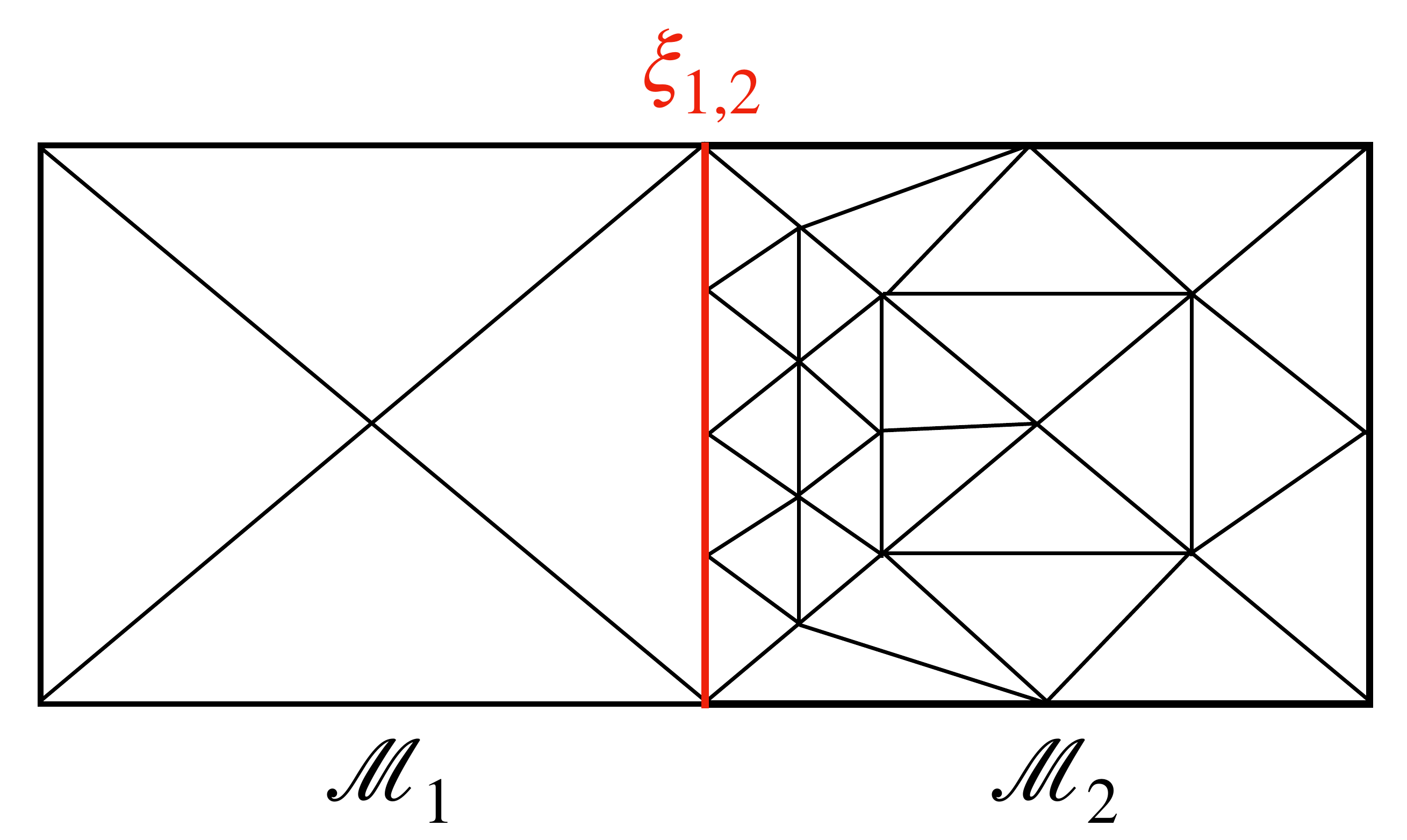} 
        } \hfill
        \parbox{0.48\linewidth}{\centering
            \includegraphics[width=0.54\linewidth]{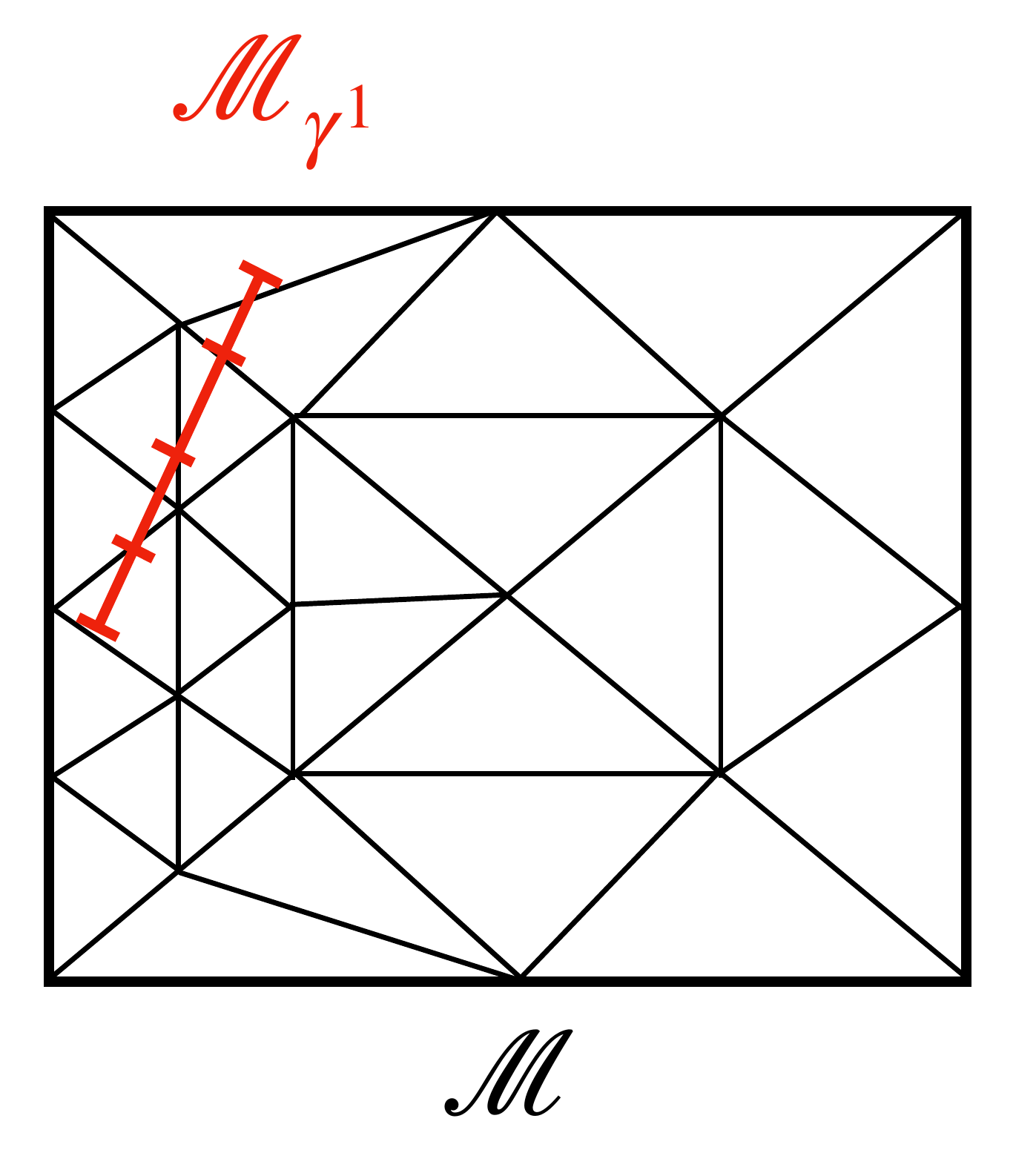} 
        }
    } \\[2em]
    \parbox{0.8\linewidth}{
        \parbox{0.48\linewidth}{(a) Non conforming interface $\xi_{1,2}$ between mesh $\mathcal{M}_1$ and mesh $\mathcal{M}_2$. } \hfill
        \parbox{0.48\linewidth}{(b) Meshes for fracture $\mathcal{M}_{\gamma^1}$ and porous matrix $\mathcal{M}$ for the embedded scenario. }
    }
    \caption{Non-conformity with matching geometry (a) and non-matching geometry (b).}
    \label{fig:comparedisc}
\end{figure}

The presented methods handles non-conforming meshes in two stages. 
The first stage involves the domain decomposition of the porous matrix. For instance, a matrix domain $\Omega$ can be split into the two domains $\Omega_1$ and $\Omega_2$ with interface $\xi_{1,2}$, this interface could be interpreted as a fracture. 
However, in discrete settings (Fig.~\ref{fig:comparedisc} (a)) it could also be a convenient way to handle different resolutions for the meshes $\mathcal{M}_1$ and $\mathcal{M}_2$, and ensure the continuity of the solution at $\xi_{1, 2}$ using the standard mortar approach introduced in~\cite{bernardi1993domain}. In practice, we need to assign the standard master and slave role, for instance, we assign the master role to $\xi_{1}$ and slave role to $\xi_{2}$.
Once the transfer operator is assembled the porous-medium-matrix system is condensed. Note that in the resulting system of equations the degrees of freedom associated with the slave discretization are eliminated after static condensation (e.g., by replacing the related rows of the matrix with the identity and the related right-hand side value with zero). Once this is achieved we go to the next stage. 

The second stage involves the embedded case (Figure~\ref{fig:comparedisc}(b)) where we compute the transfer operator between $W_h$ and $W_{h,\gamma}$ and condense the resulting system as described in Section~\ref{sec:disc}.

\subsection{Information transfer} \label{sec:transfer}
The computation of the so called mortar integrals, which are the integral terms in~\eqref{eq:matsys}, \eqref{eq:fracsys}, and \eqref{eq:lagrsys}, associated with the Lagrange multiplier require special handling. In fact, quadrature formulas have to be generated in the intersection between elements of the matrix and the fracture.
The computation of intersections differs depending on which type of fracture is considered. Table~\ref{tab:isectype} lists, for each type of coupling, the roles and the intersection types. Here, the master role is given to the domain covering completely the slave domain. The slave role is given to the domain with which we associate the Lagrange multiplier space.

For the coupling at an interface $\xi$, either for non-conforming domain decomposition or for an interface fracture $\gamma_\xi$, the coupling is performed on a common surface description. This operation requires intersecting oriented planar polygonal elements in the case of a three-dimensional problem, or intersecting oriented line elements in the case of two-dimensional problem.

For the embedded scenario the polytopal element of the matrix are intersected with the lower- or equi- dimensional polytopal elements of the fracture.
The case where we have $\mathcal{M} \subset \mathbb{R}^d$ and $\mathcal{M}_\gamma$ being a $(d-1)$-dimensional manifold mesh, requires particular handling when the fracture elements are aligned with the surface of the matrix elements. In such case it is likely that some intersections might be computed twice, hence these duplicate intersection are detected and removed. 

The $(d-1)$-dimensional  dimensional fractures represented with the mortar method require a careful set-up. 
In fact, in the case of intersecting fractures, multiple sub-domains (more than two) might intersect at one point or edge.
The mesh primitives, i.e. edges and nodes, generating such intersections require a specific handling.  
The first approach consists of ignoring the side elements that are incident to the aforementioned mesh primitives when defining the discrete Lagrange multiplier as in~\citet{krause2015}, hence introducing discontinuities of the solution at these interfaces. 
The second approach, which would require a set-up similar to~\citet{farah2018}, consists of defining one-to-many relationships for the intersecting primitives.  Here, one master primitive has to be determined and continuity is either enforced using interpolation for intersecting nodes, or with a weak equality condition for intersecting edges.
The third approach, consists of explicitly defining the entire surface mesh for which the discrete Lagrange multiplier is constructed.
These complications with the mortar method are not present in the equi-dimensional case if the discrete fractures are represented with a unique conforming mesh, since the role of master and slave can be trivially assigned to porous-matrix and fracture respectively.

The computation of the intersections listed in Table~\ref{tab:isectype} is performed with suitable variants of the Sutherland-Hodgman clipping algorithm~\cite{sutherland1974}.
Once the intersection is computed, if required, this intersection is meshed into a simplicial complex so that we can map quadrature rules to each simplex and integrate exactly. 

\begin{table}[]
    \centering
    \begin{tabular}{ccl}
    \hline
        \textbf{Matrix (master)} & \textbf{Fracture (slave)} & \textbf{Intersection type} \\
        \hline
        $\Omega \subset \mathbb{R}^3$ & $\gamma^3$ & polyhedron-polyhedron \\
        $\Omega \subset \mathbb{R}^3$ & $\gamma^2$ & polyhedron-polygon \\
        $\partial\Omega_i \cap \xi_{i,j}  \subset \mathbb{R}^3$ & $\partial\Omega_j \cap \xi_{i,j}  \subset \mathbb{R}^3$ & polygon-polygon (oriented) \\
        $\Omega \subset \mathbb{R}^2$ & $\gamma^2$ & polygon-polygon \\
        $\Omega \subset \mathbb{R}^2$ & $\gamma^1$ & polygon-segment \\
        $\partial\Omega_i \cap \xi_{i,j} \subset \mathbb{R}^2$ & $\partial\Omega_j \cap \xi_{i,j}  \subset \mathbb{R}^2$ & segment-segment (oriented)\\
        \hline
        \vspace{1em}
    \end{tabular}
    \caption{Intersection for different coupling types. The standard master and slave roles used in the mortar literature are associated with matrix and fracture, respectively.}
    \label{tab:isectype}
\end{table}

\subsection{Adaptive refinement} \label{sec:refine}
\begin{figure}[ht]\centering\footnotesize
    \includegraphics[width=0.4\linewidth]{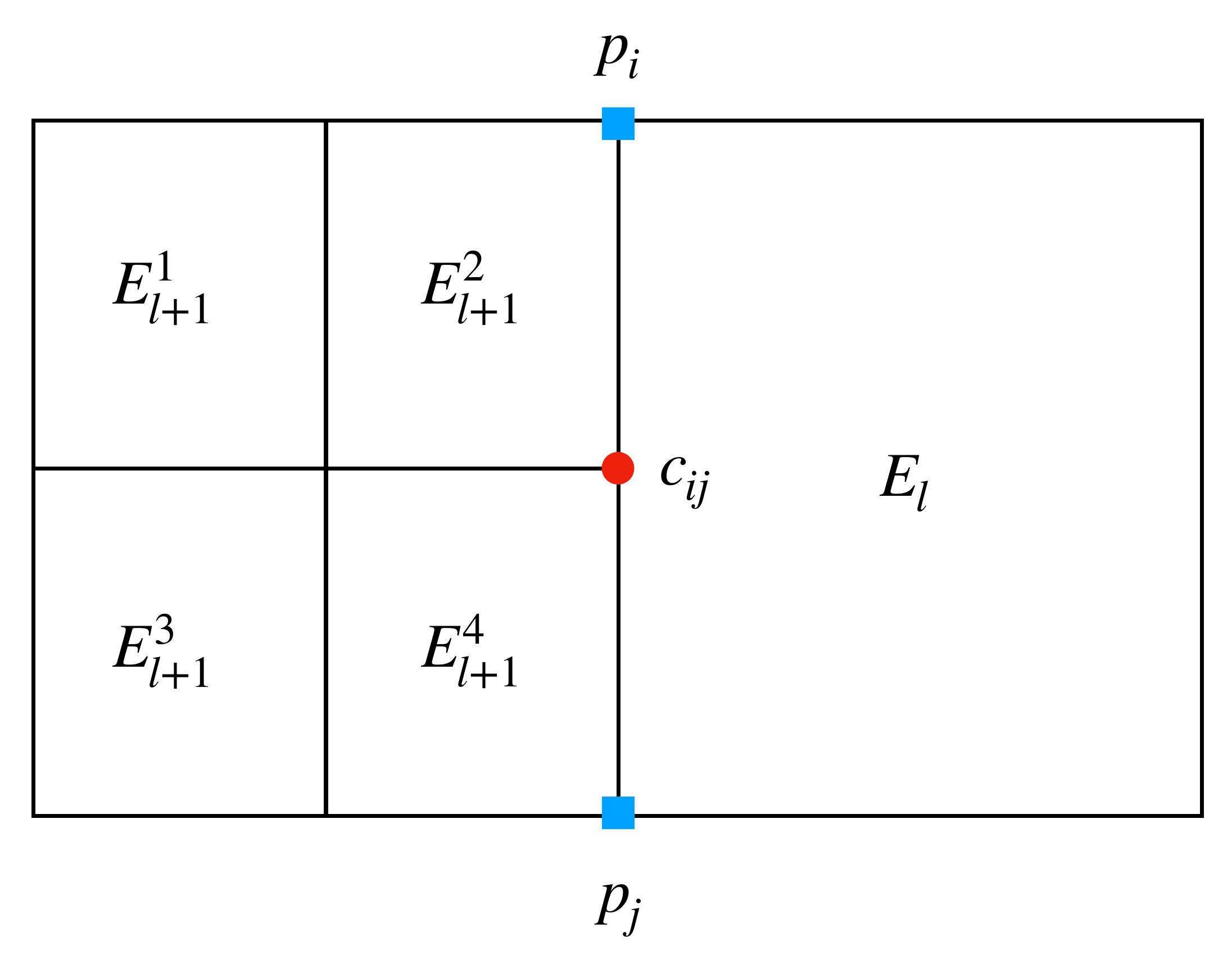}
    \caption{%
    Bi-linear elements $E^k$ with different levels of refinement $k \in \{l, l+1\}$. 
    With $p_i, p_j$ we denote the parent nodes that are generated on level $l$, with $c_{ij} = (p_i+p_j) / 2$ we denote the child node generated on level $l+1$ by splitting the edge $(i,j)$. 
    Depending on the finite element discretization the splitting is done in different (multiple) locations.
    The prolongation operation at node $c_{ij}$ for this particular case would simply be $u(c_{ij}) = (u(p_i)+u(p_j)) / 2$.}
    \label{fig:hangingnode}
\end{figure}

Let us recall the definition of element $E$ from Section~\ref{sec:disc}. With $\partial E$ we denote the boundary of element $E$ and with $\bar{E}$ its closure.
A mesh is said to be conforming if $\bar{E}_i \cap \bar{E}_j, i \neq j$ is a common vertex, edge, face, or $\emptyset$. A node is said to be hanging if it lies on the interior of an edge or face of another element. A mesh containing at least one hanging node is called non-conforming.

Non-conforming adaptive mesh refinement has the advantage that can be applied to any type of element in a rather straight-forward and independent manner. 
However, once the elements marked by the error estimator are refined, the resulting mesh might have hanging nodes, as shown in Figure~\ref{fig:hangingnode}. 

As a consequence, continuity of the solution is to be enforced either by employing variational restriction or discontinuos Galerkin methods. Here, we consider the variational restriction technique which is thoroughly explained in~\citet{erven2019nonconforming} also for high-order discretizations. 

Let $\mathbf{R}_i, i \in \{m, s\}$ be a suitable restriction operator that splits contributions of each hanging node to its adjacent nodes, where $m$ stands for master and $s$ for slave.
For combining adaptivity with DFMs and the method of Lagrange multipliers, we consider the constrained spaces arising from the refinement and variational restriction, which requires us to perform some slight modifications to the final steps of the assembly procedure for the transfer operator, in any of the cases we mentioned in previous sections. We recall the definitions of the coupling matrix $\mathbf{B}$ and mass-matrix $\mathbf{D}$ from Section~\ref{sec:disc}, and define the modified transfer operator
$$
\mathbf{T}_R = (\mathbf{R}_s^T \mathbf{D} \mathbf{R}_s)^{-1} (\mathbf{R}_s^T \mathbf{B} \mathbf{R}_m).
$$
The operator $\mathbf{T}_R$ allows us to transfer between constrained spaces, however for also setting the values in the hanging nodes we apply the prolongation operator $\mathbf{P}_s = \mathbf{R}_s^T$ as follows
$$
\mathbf{T} = \mathbf{P}_s \mathbf{T}_R.
$$
This small modification allows to use $\mathbf{T}$ as in the standard case without any special treatment as described in Section~\ref{sec:varform}.

%
\subsection{Implementation} \label{sec:implementation}
The routines described in this paper are implemented within the open-source software library \emph{Utopia}~\cite{utopiagit}. In this work, \emph{Utopia} uses \emph{libMesh}~\cite{libmesh} for the finite element discretization, \emph{MOONoLith}~\cite{moonolith} for the intersection detection, and \emph{PETSc}~\cite{petsc} for the linear algebra calculations. 
The software developed for this contribution is used by means of a JSON (JavaScript Object Notation) input file where any number of mesh files can be linked to the simulation and coupled together automatically for creating complex networks of fractures.

%
\section{Numerical results \& discussion} \label{sec:results}

First, the focus is on a comparison of results obtained with the equi-dimensional embedded technique and the mortar  method to a specific selection of commonly used 2D and 3D benchmarks~\cite{flemisch2018,berre_2020}.
Here, we use non-conforming mesh refinement in proximity of the fractures for maintaining the size of the mesh small while achieving small deviations from the reference solution.
Second, we show how employing an adaptive mesh refinement, with gradient-recovery based error estimator~\cite{yan2001gradient}, allows us to refine only where the error in the solution is estimated to be higher.
Finally, we present a complex 3D experiment inspired by realistic scenarios where we conveniently mix all the techniques we covered in this article.
In each of the following sections we discuss the practicalities of the different techniques and how to combine them.

For compactness, in the following sections we use the abbreviation ED for equi-dimensional and HD for hybrid-dimensional. 
We report exclusively the error of the solution associated with the matrix discretization. This is done for avoiding redundant information, 
since the solution for the fracture is just the $L^2$-projection of the solution for the matrix.

\subsection{2D Benchmarks}
The 2D settings allows us to provide a simpler and clearer presentation of the numerical results. 
Hence, with a selection of 2D benchmarks from \citet{flemisch2018}, we complement and extend the previous contribution presented in~\citet{SCHADLE201942}. 
We verify the embedded ED approach and our implementation of the mortar method with the dual Lagrange multiplier. 
We show how adaptive mesh refinement allows us to solve problems with a smaller number of degrees of freedom while achieving the desired accuracy in the solution. For all 2D cases we use the reference solution proposed in~\citet{flemisch2018} which is computed the mimetic finite difference method.


\subsubsection{2D Benchmark Case 1: Regular fracture network}\label{sec:regular}


\begin{figure}[ht] \centering\footnotesize
    \parbox{0.8\linewidth} {
    \includegraphics[width=0.49\linewidth]{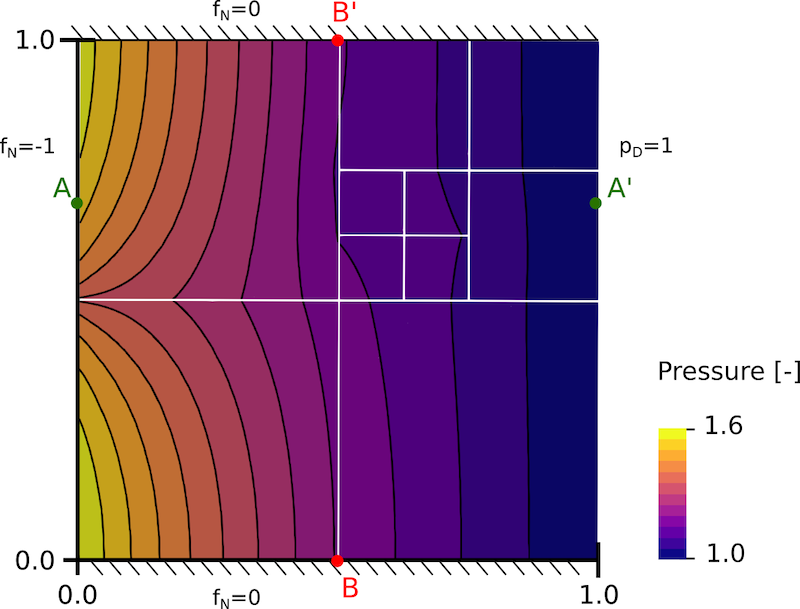} \hfill
    \includegraphics[width=0.49\linewidth]{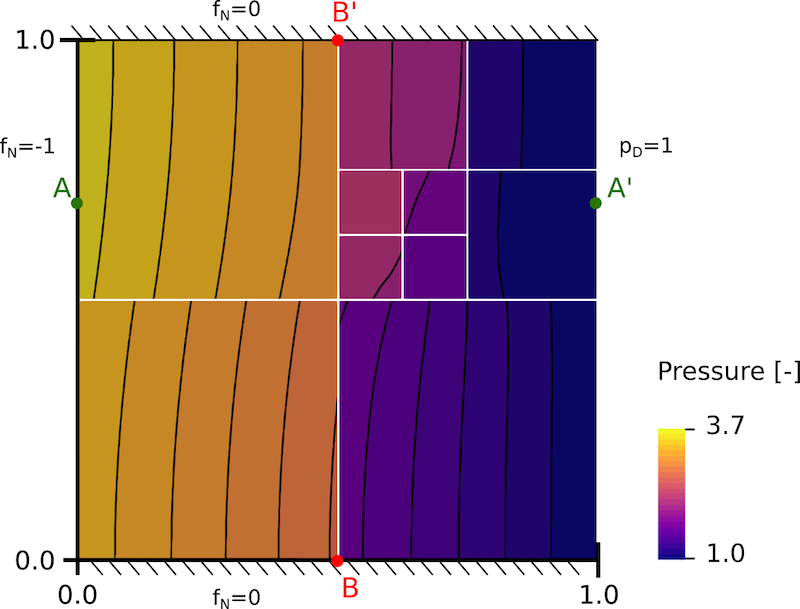} 
    } \\[2em]
    \parbox{0.8\linewidth}{
        \parbox{0.45\linewidth}{\centering (a) Embedded/conductive} \hfill
        \parbox{0.45\linewidth}{\centering (b) Mortar/blocking }
    }
    \caption{%
    \emph{Benchmark 1} in~\citet{flemisch2018}. Pressure solution for regular fracture network with six equi-dimensional fractures for
    conductive fractures with the equi-dimensional embedded method (a) and blocking fractures with the equi-dimensional mortar method (b).
    }
    \label{fig:regularresults}
\end{figure}


 \begin{figure}[ht]
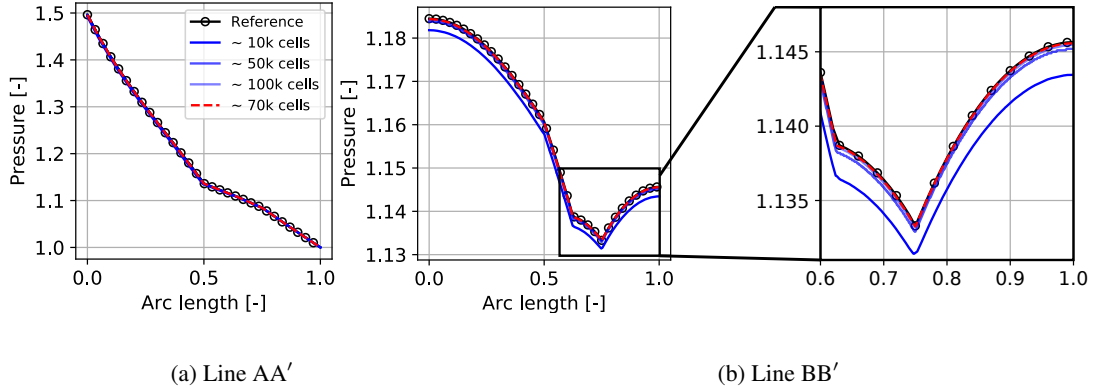
 \centering\footnotesize
    \parbox{0.8\linewidth} {
    \includegraphics[width=0.328\linewidth]{figures/regualar_embedded_mortar_plot1.pdf} \hfill
    \includegraphics[width=0.765\linewidth]{figures/regualar_embedded_mortar_plot2.pdf}
    } \\[2em]
    \parbox{0.8\linewidth}{
        \parbox{0.45\linewidth}{\centering (a) Line \aaline{}} \hfill
        \parbox{0.45\linewidth}{\centering (b) Line \bbline{}}
    }
    \caption{%
    \emph{Benchmark 1} in~\citet{flemisch2018}: Pressure profiles along the lines \aaline{} (a) and \bbline{} (b) with a zoom into the area with the largest deviation. Blue lines indicate Emebedded-ED,  dashed red lines indicate Mortar-ED.
    }
    \label{fig:regularmortar}
\end{figure}


\begin{table}[ht]
    \caption{%
    \emph{Benchmark case 1} in~\citet{flemisch2018}. For each method we report number of elements in the matrix (\textbf{\#-matr}) and fracture (\textbf{\#-frac}), number of degrees of freedom (\textbf{d.o.f.}), normalized number of non zero entries (\textbf{nnz/size{$^2$}}), condition number ($\| \cdot \|_2$-\textbf{cond}.) and error in the matrix (\textbf{err}$_m$) with respect to the reference solution.
}
    \input{\tab results_2DRegular.tex}
    \label{tab:regular2D}
\end{table}


We consider the same settings and reference solutions used in~\citet{flemisch2018} \emph{Benchmark 1}, with both conductive and blocking fractures, as shown in Fig.~\ref{fig:regularresults}.
Both settings have the same square domain $\Omega = [0, 1]^2 $ and boundary conditions. We impose Dirichlet conditions on the right boundary,  where the pressure is set to constant value 1. We impose no-flow conditions on the bottom and top sides. Permeability of the matrix is uniform 
$\mathbf{K} = \mathbf{I}$, where $\mathbf{I}$ is the identity matrix. We distinguish conductive and blocking scenarios for the fracture permeability. For the conductive scenario, in the HD case the permeability tensor is described as 
$\mathbf{K}_{\gamma^1} = \epsilon \mathbf{I}  \cdot  10^4 $, where $\epsilon = 10^{-4}$ is the fracture aperture,
whereas for the ED case we have $\mathbf{K}_{\gamma^2} = \mathbf{I} \cdot 10^4$. For the blocking scenario, we only have the ED case with $\mathbf{K}_{\gamma^2} = \mathbf{I} \cdot 10^{-4}$.

A particular emphasis is placed on equi-dimensional fractures and the comparison between non-conforming embedded/immersed fractures and geometrically conforming fractures which are glued together with the matrix using the mortar method. For the hybrid-dimensional embedded case we consider the results of~\citet{SCHADLE201942} using the methods of dual-Lagrange multipliers and static condensation.
In Table~\ref{tab:regular2D}, we report results for three different resolutions for each of the three strategies. 
In the following paragraphs we illustrate the different set-ups, results, and limitations of the embedded and mortar methodologies.

The employed embedded techniques enforce the continuity of the solution at the intersection of the matrix and fracture meshes. 
Hence, steep pressure jumps and barriers can not be represented. Consequently, for the embedded case we restrict our study to conductive fractures.
In particular, we study the equi-dimensional embedded technique for 3, 5, and 6 levels of non-conforming mesh refinement (Section~\ref{sec:refine}) performed exclusively in proximity of the fractures.
This refinement patterns are generated automatically using the variational transfer algorithm for marking the elements of the matrix that are intersecting with the fractures.
On Fig.~\ref{fig:regularmortar}, we plot the solution over the lines \aaline{} (y = 0.75) and \bbline{} (x = 0.5) and it can be observed that even for low resolutions, the solutions are in agreement with the reference results of~\citet{flemisch2018}.
From Table~\ref{tab:regular2D}, it can be observed that with this particular set-up we reach an error in  the order of $10^{-7}$. Note that the number of elements in the fracture network do not influence the number of degrees of freedom, hence they do not count for the computational cost of solving the linear system but only for the set-up phase which is typically cheaper.

For the mortar method based experiment we are required to represent the fracture explicitly in the  matrix mesh. For this particular scenario,  the fracture is modelled as an equi-dimensional geometry which is meshed independently from the matrix. 
This allows us to refine all the different sub-domains in a completely independent manner and glue them together using the mortar method. 
We manually refine the matrix around the two fractures crossing at the center of the domain. This is done for having a higher resolution in the region of interest of the benchmark.
From Table~\ref{tab:regular2D}, it can be observed that even for low mesh resolutions of the matrix the error reaches an order of $10^{-8}$ for both, conductive and blocking fractures. However, the mesh resolution of the fractures is very high from the start. 
Despite this fact, in our experiments the actual degrees of freedom are only associated with one layer of nodes in the middle of the fracture, while the ones at the interface are eliminated by means of the mortar constraints as mention in Section~\ref{sec:multimesh}. 

We can observe that the embedded methodologies generate linear systems with comparable condition numbers and number of degrees of freedom. Whereas, the mortar-ED method gives rise to larger systems with larger condition numbers. 
Only with the mortar-ED we are able to solve the blocking scenario~\ref{fig:regularresults}(b), however already for this simple experiment the mesh set-up is more laborious due to the matching geometry.


\subsubsection{2D Benchmark Case 2: Hydrocoin}\label{sec:hydrocoin}
\begin{figure}[ht]\centering\footnotesize
 \centering
 \includegraphics[width=0.70\linewidth]{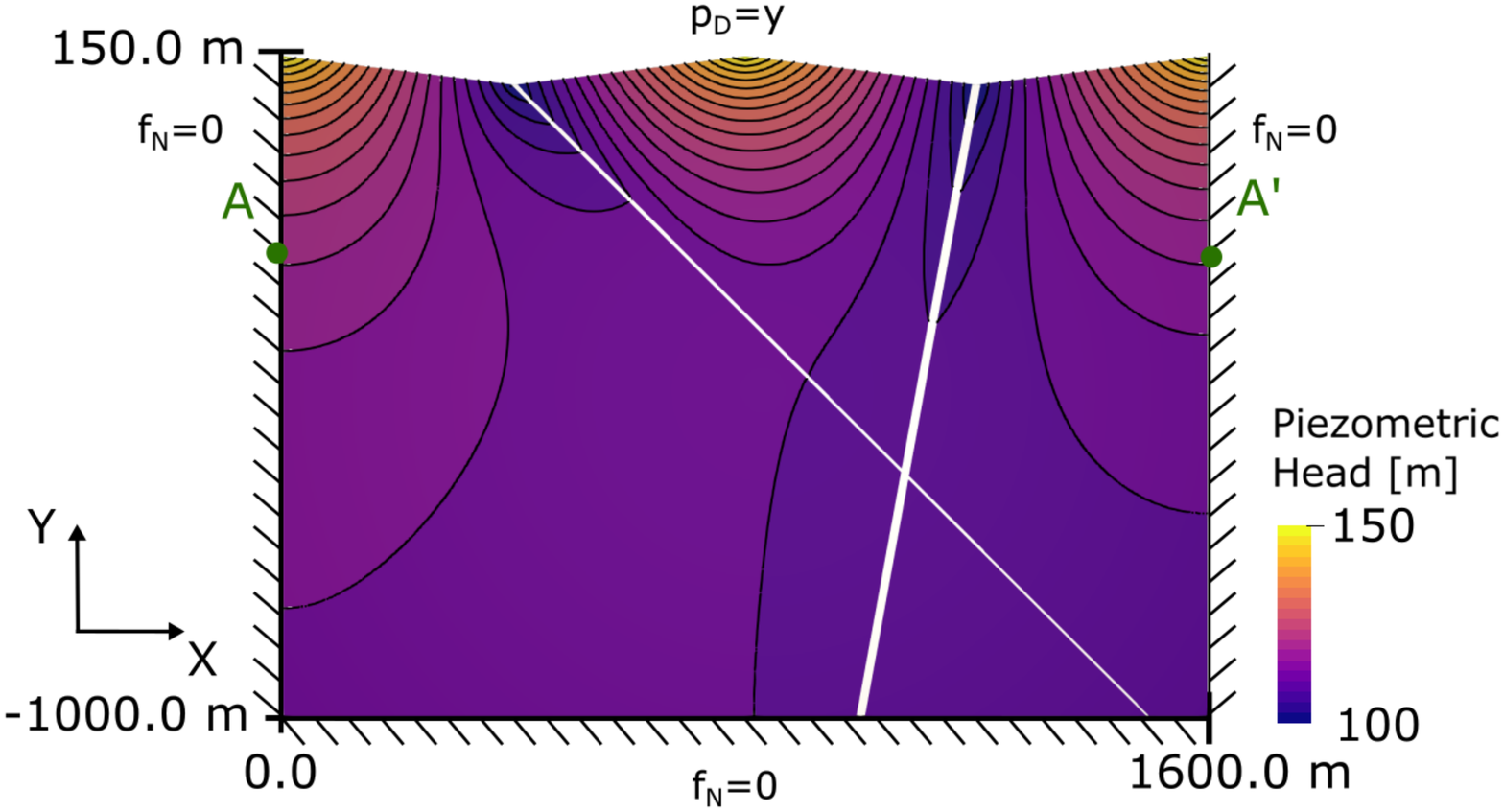}
 \caption{%
\emph{Benchmark 2} in \citet{flemisch2018}.
 Embedded-ED: spatial distribution of the piezometric head~$\text{[m]}$. Solution profiles are compared along the line \aaline{} with coordinates $y=-200\,\text{[m]}$ (marked in green).
 }
\label{fig:hydro_embedded_bc}
\end{figure}
 
\begin{figure}[ht]\centering\footnotesize
  \parbox{0.8\linewidth} {
 \includegraphics[width=0.31\linewidth]{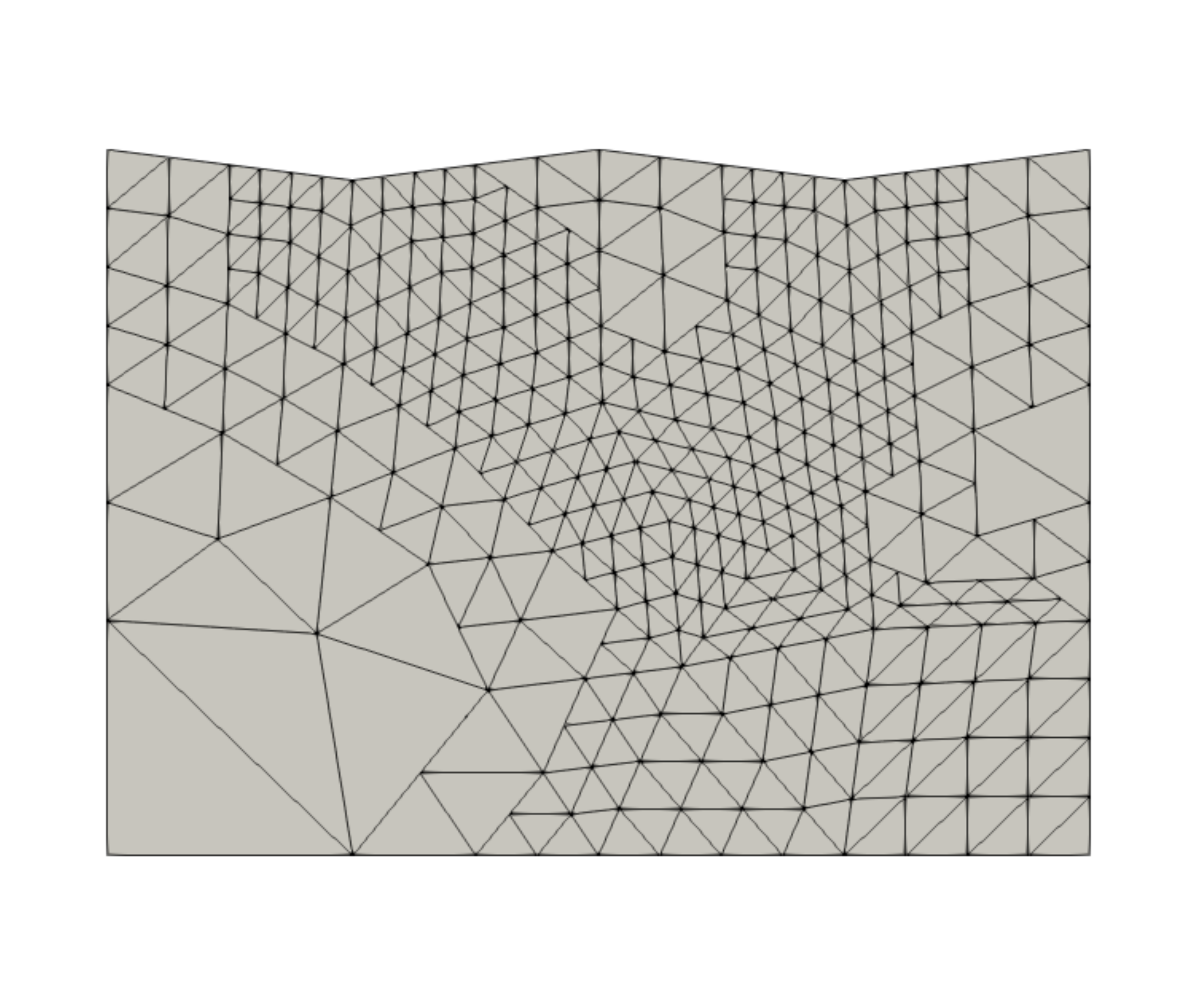}\hfill
  \includegraphics[width=0.30\linewidth]{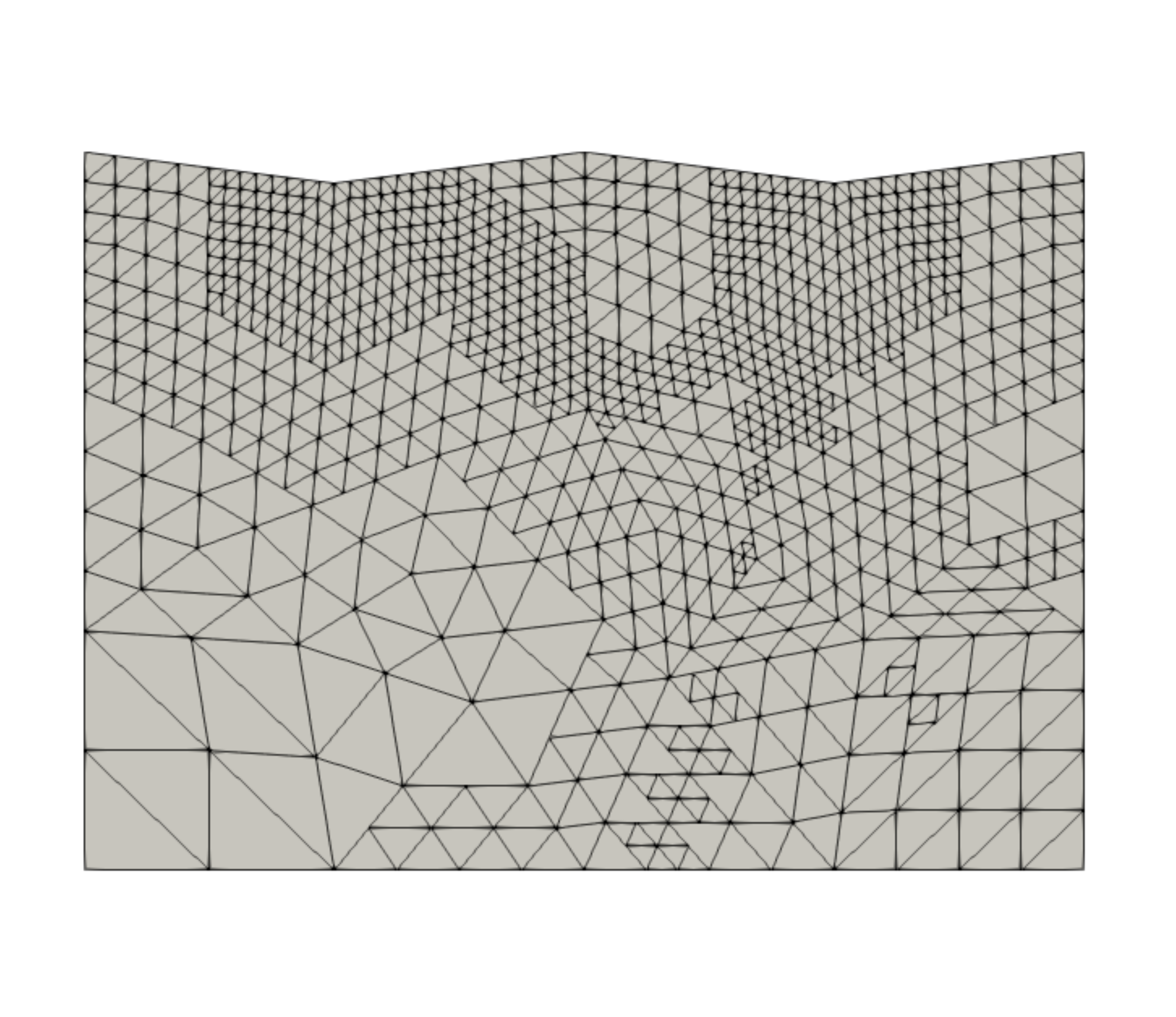}\hfill
  \includegraphics[width=0.30\linewidth]{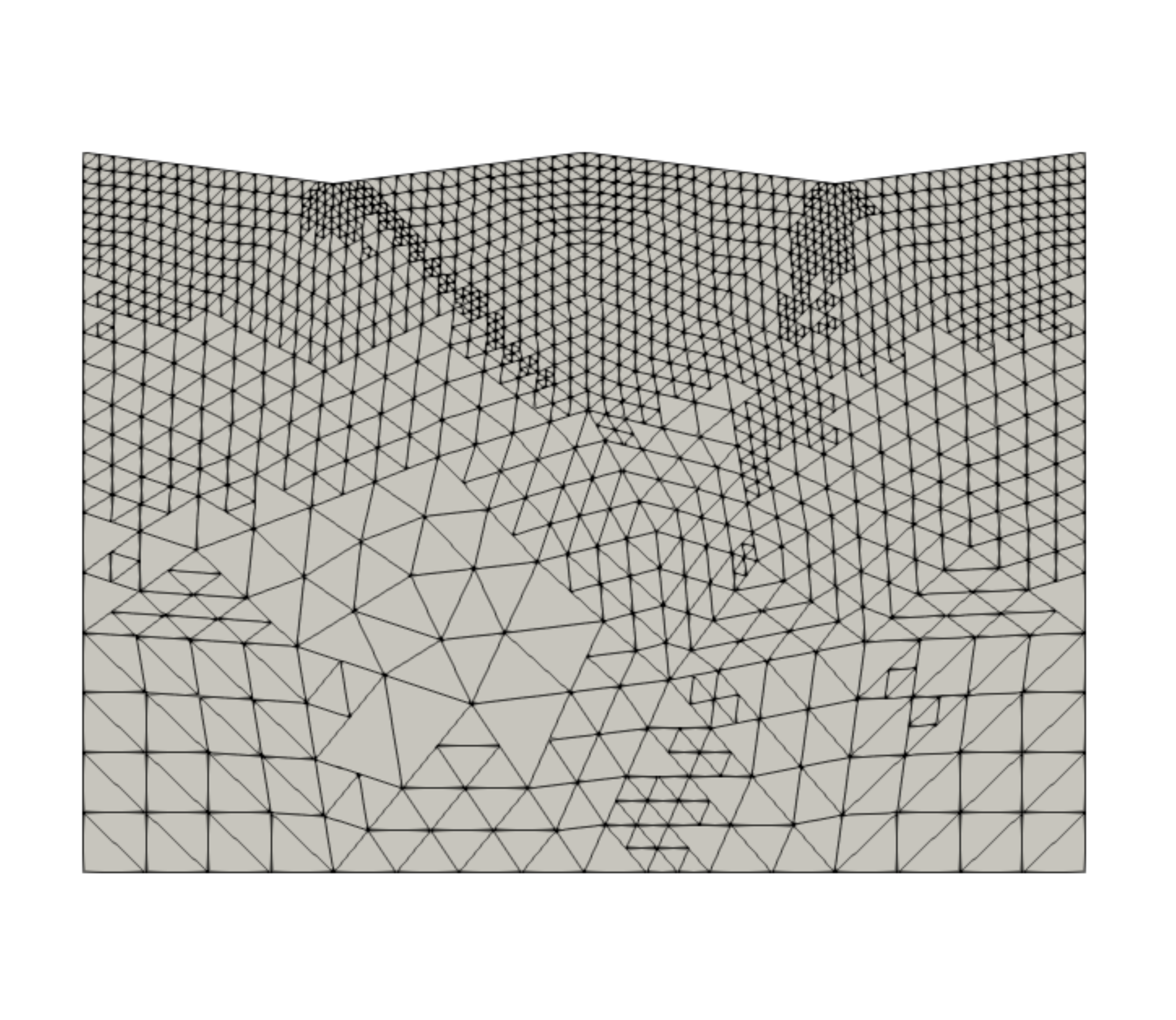}\hfill
  }
\parbox{0.8\linewidth}{
        \parbox{0.30\linewidth}
        {\centering 1 AR} \hfill
        \parbox{0.30\linewidth}
        {\centering 2 AR} \hfill
       \parbox{0.30\linewidth}
        {\centering 3 AR} \hfill
    }
  \parbox{0.8\linewidth} {\includegraphics[width=0.31\linewidth]{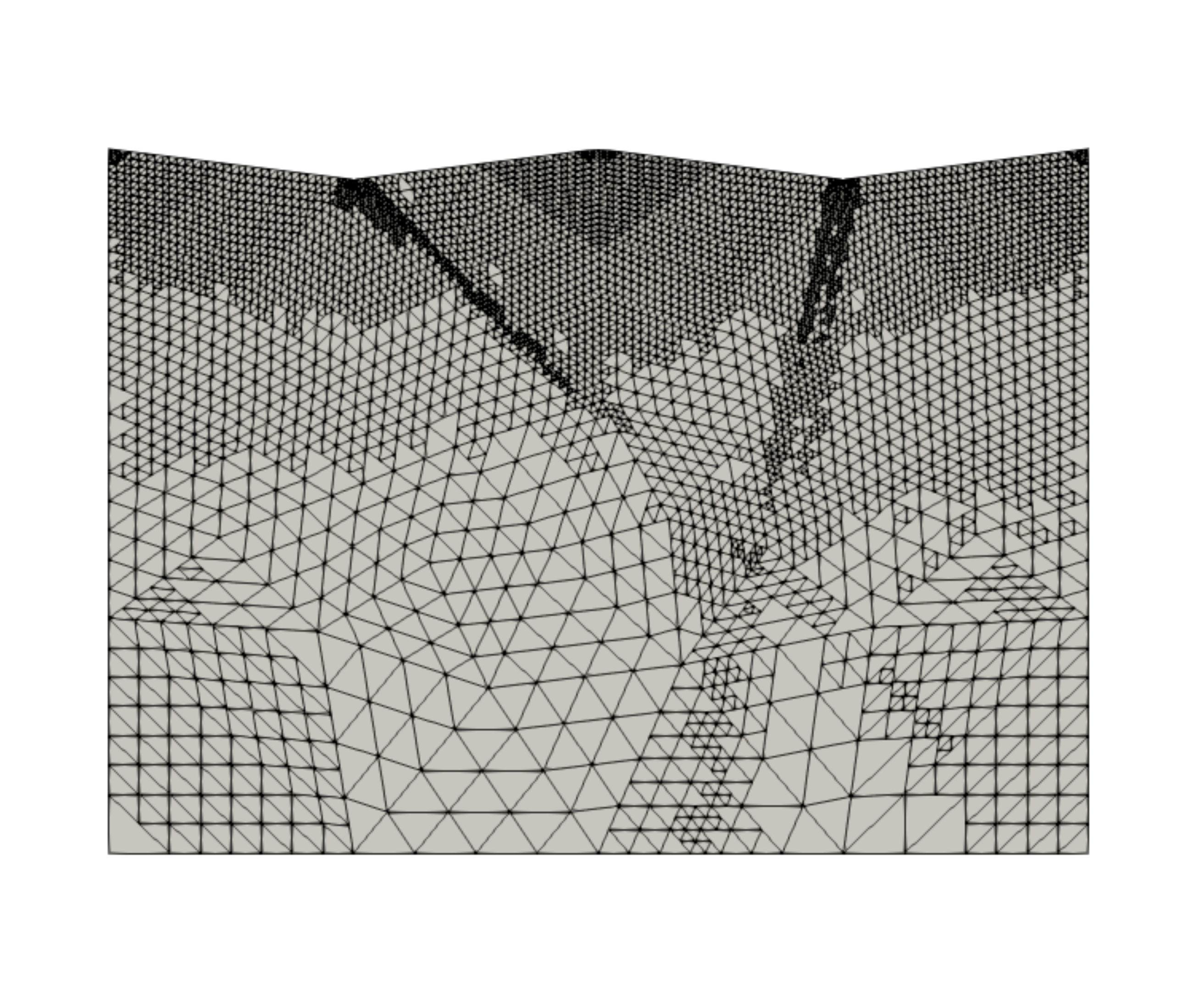}\hfill
\includegraphics[width=0.30\linewidth]{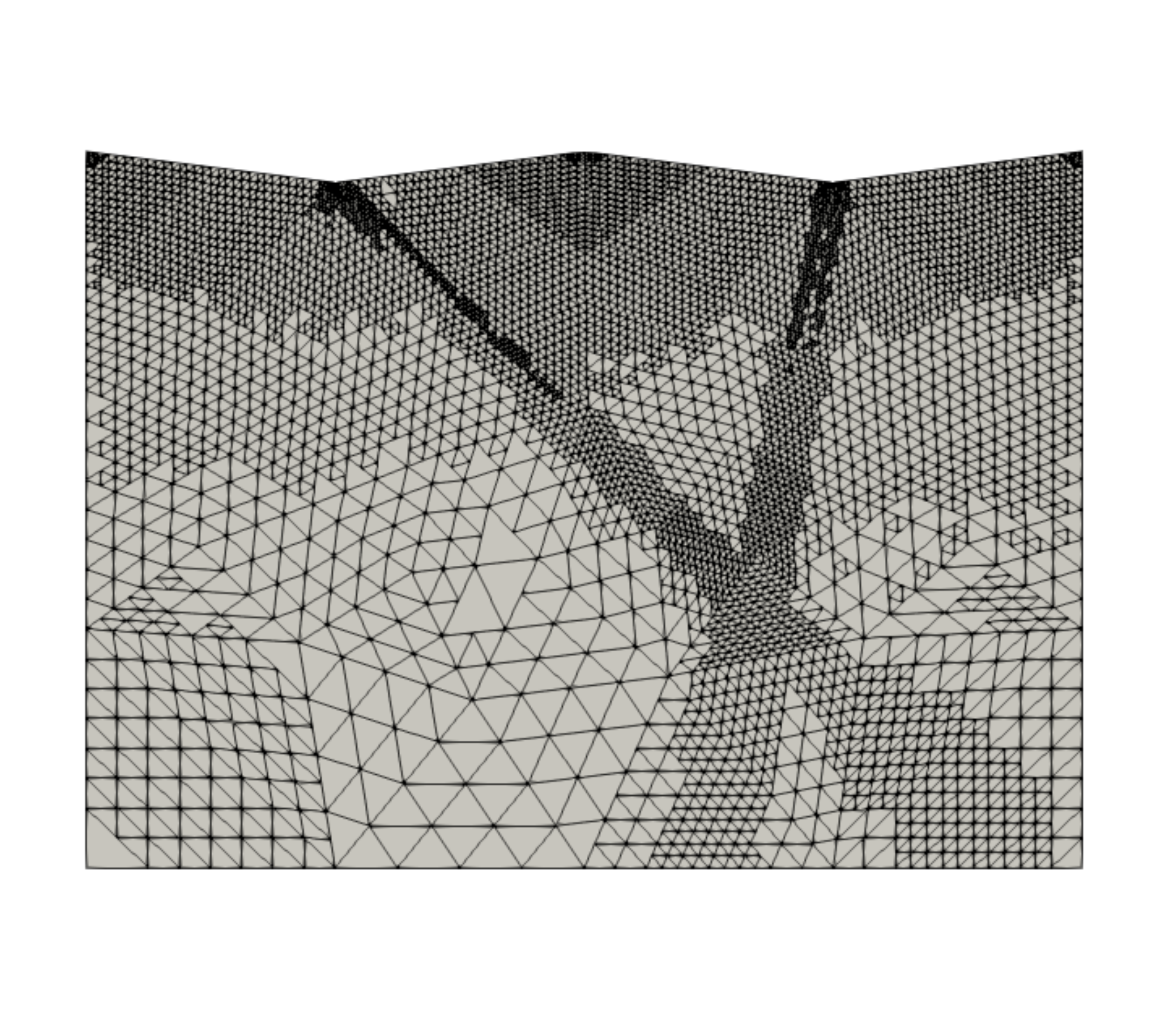}\hfill
\includegraphics[width=0.30\linewidth]{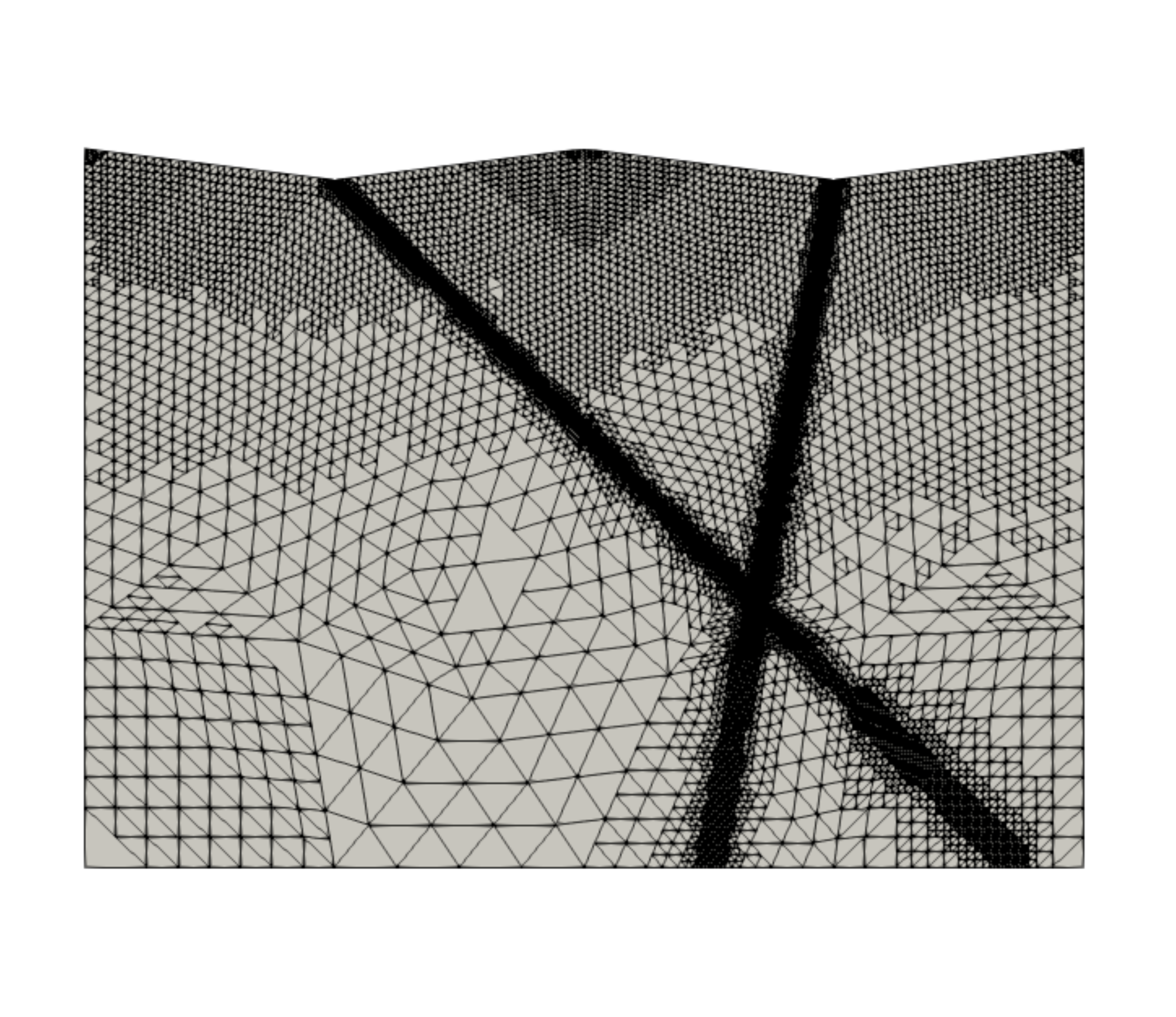}}
    \parbox{0.8\linewidth}{
        \parbox{0.30\linewidth}
        {\centering 4 AR} \hfill
        \parbox{0.30\linewidth}
        {\centering 5 AR} \hfill
        \parbox{0.30\linewidth}
        {\centering 6 AR }
    }
 \caption{Embedded-ED adaptive mesh refinement. The meshes have been obtained by performing first an adaptive refinement in the region where fractures are located, then by using the gradient-recovery based error estimator.}
\label{fig:hydro_embedded_mesh_amr}
 \end{figure}
 
\begin{table}
     \caption{%
        Uniform refinement for the Embedded-ED vs Mortar-ED cases. 
        Here $\textbf{err}_m$ is the error computed with respect to the reference solution.
        $\#\textbf{UR}$ refers to the number of uniform refinements.
     }
     \input{\tab results_2DHydroCoinUR.tex}
     \label{tab:tableHydroEmbeddedU}
\end{table}

 \begin{figure}[ht]\centering\footnotesize
      \parbox{0.8\linewidth} {
 \includegraphics[width=0.49\linewidth]{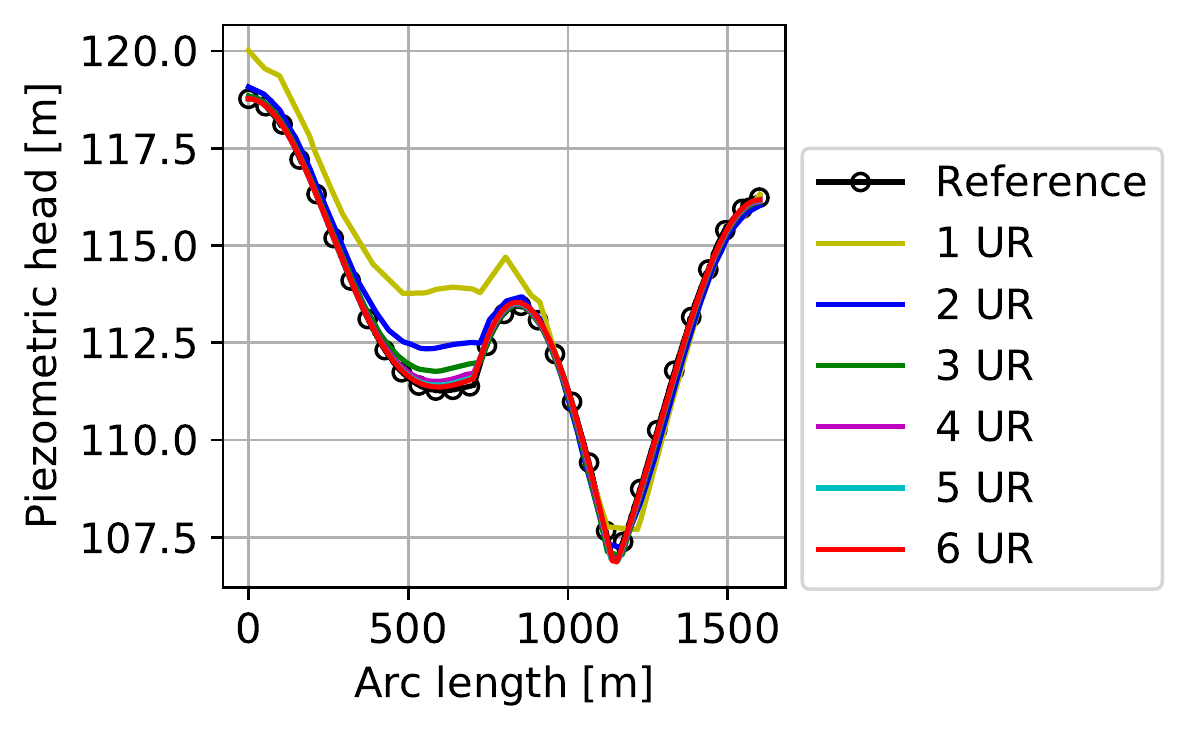} \hfill
     \includegraphics[width=0.49\linewidth]{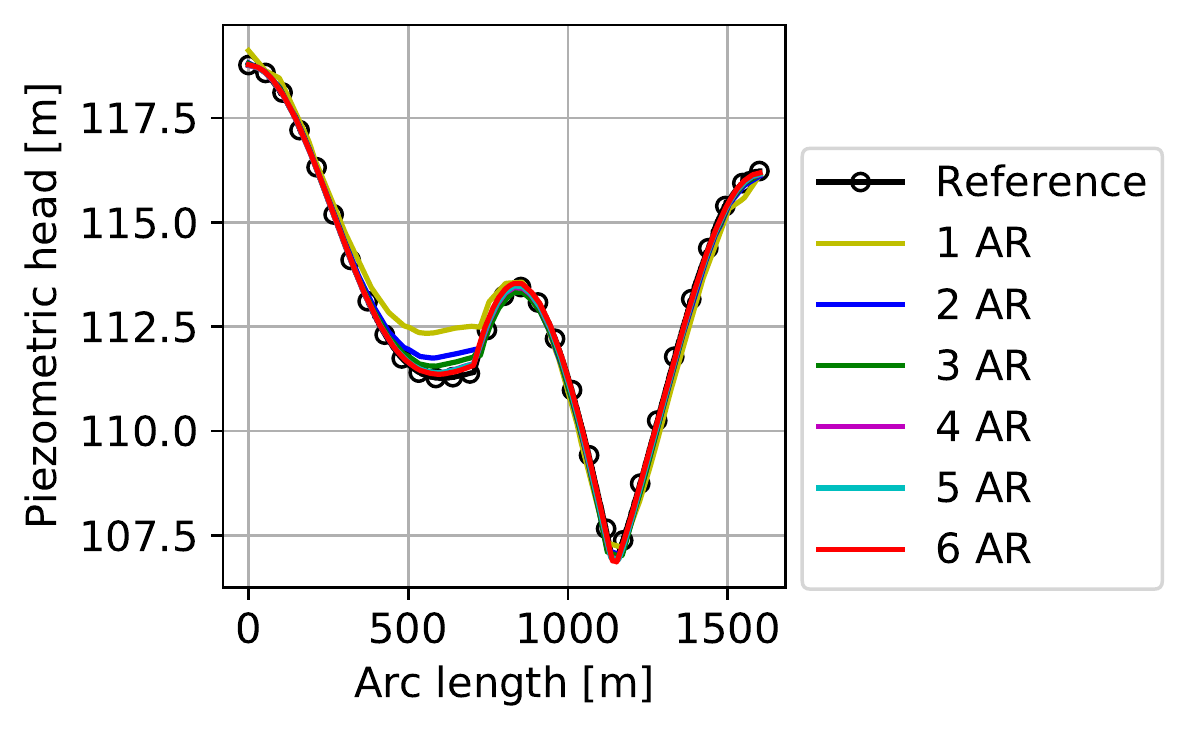}
    } \\[2em]
    \parbox{0.8\linewidth}{
        \parbox{0.45\linewidth}{\centering (a) Uniform Refinement (UR)} \hfill
        \parbox{0.45\linewidth}{\centering (b) Adaptive Refinement (AR) }
    }
   \caption{%
    Pressure along line \aaline{} for the uniform (A) and the adaptive (B) Embedded-ED test cases. Here, $\text{UR}$ and $\text{AR}$ refer to  different steps of uniform and adaptive refinements performed on the initial matrix mesh. In particular, the number of $\text{AR}$  are obtained combining a gradient recovery strategy with an adaptive refinement performed on the overlapping region between the matrix and the fracture zone. More details can be found in  tables~\ref{tab:tableHydroEmbeddedA} and ~\ref{tab:tableHydroEmbeddedU}. 
      }
\label{fig:hydro_embedded_ED}
 \end{figure}
 
 \begin{table}
    \caption{Adaptive refinement for the Embedded-ED and Mortar-ED cases. Here, $\textbf{err}_g$ is the threshold used for the gradient recovery strategy. $\#\textbf{AR}$ refers to the number of adaptive  refinements, whereas the number within parentheses refers to the steps of adaptive refinement performed on the overlapping region between the matrix and the fracture meshes.}
    \input{\tab results_2DHydroCoinAMR.tex}
    \label{tab:tableHydroEmbeddedA}
\end{table}

\begin{figure}[ht]\centering\footnotesize
     \parbox{0.8\linewidth} {
    \includegraphics[width=0.49\linewidth]{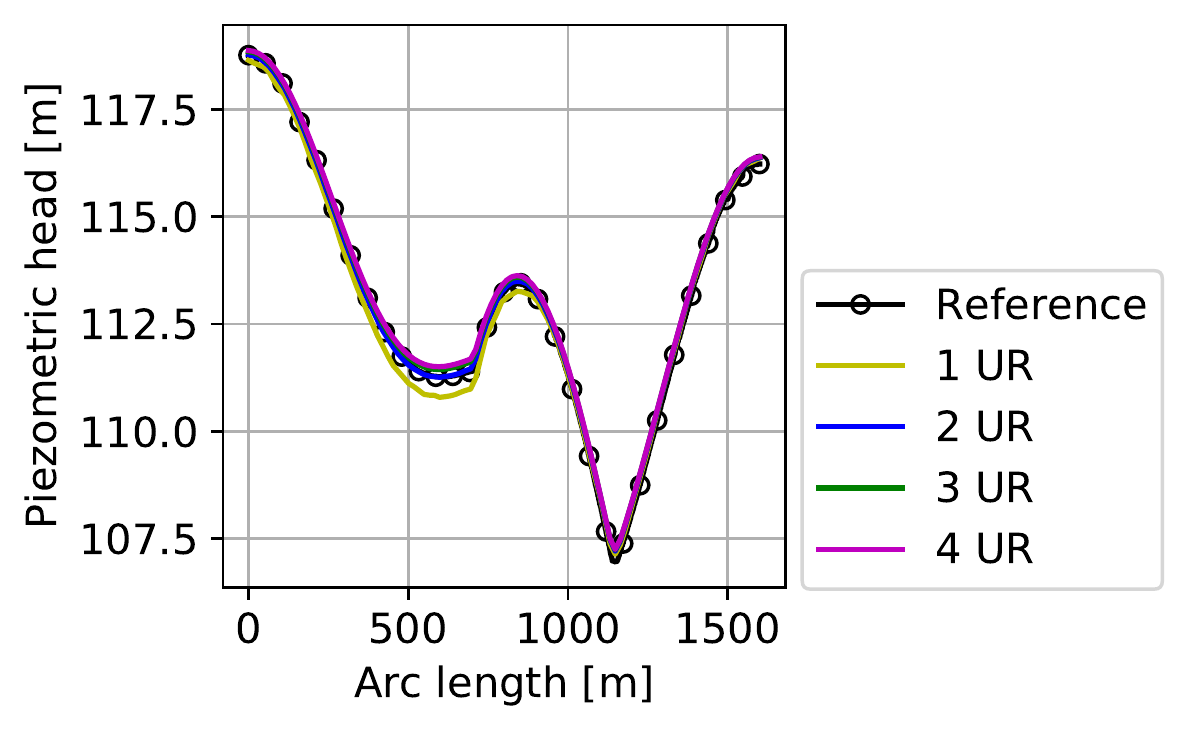} \hfill
    \includegraphics[width=0.49\linewidth]{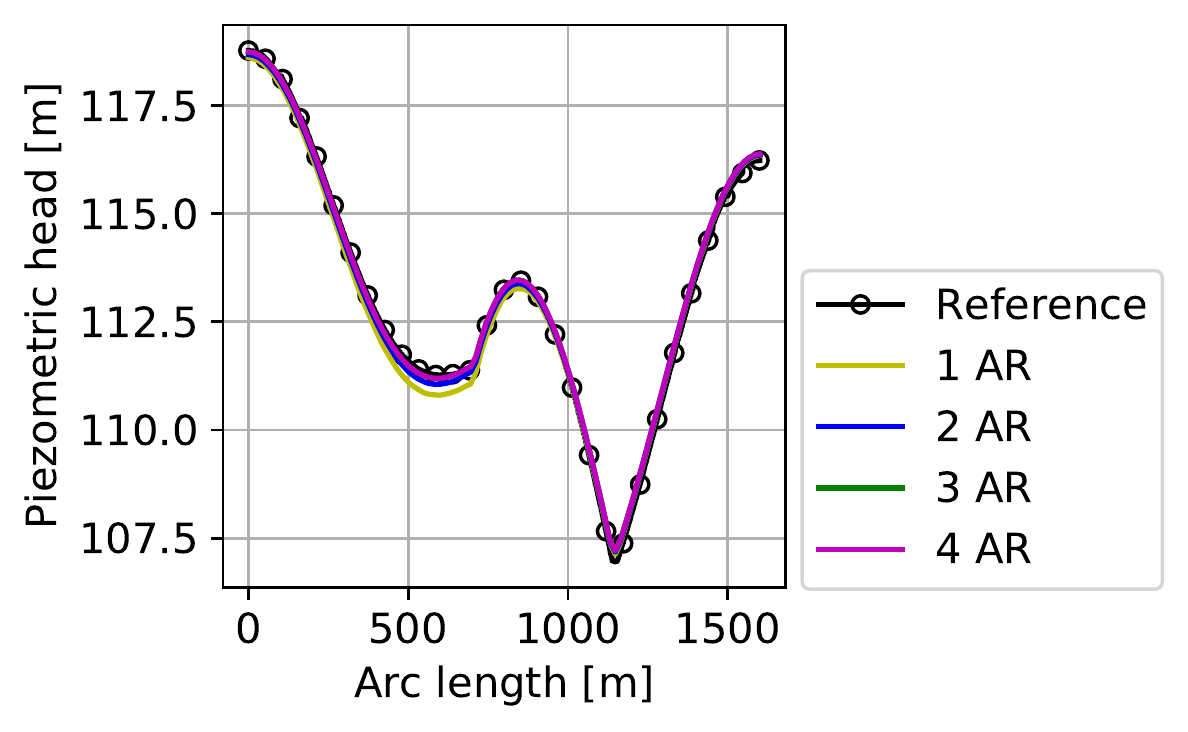} 
    } \\[2em]
    \parbox{0.8\linewidth}{
        \parbox{0.45\linewidth}{\centering (a) Uniform Refinement (UR)} \hfill
        \parbox{0.45\linewidth}{\centering (b) Adaptive Refinement (AR)}
    }
   \caption{%
  Pressure along line \aaline{} for the uniform (A) and the adaptive (B) Mortar-ED test cases. Again, $\text{UR}$ and $\text{AR}$ refer to the steps of uniform and adaptive refinement performed on the initial matrix mesh.   
 We use the MFD solution proposed in~\citet{flemisch2018} as a reference.}
 
\label{fig:hydro_embedded_MR}
 \end{figure}

We test the capability of the adaptive mesh refinement by considering the hydrocoin benchmark~\cite{nicholson1987nrc} for two different techniques: the embedded-ED and the mortar-ED. 
The matrix consists of a rectangular box with a broken line located on the top, whereas the fracture network consists of two oblique lines. As shown in Fig.~\ref{fig:hydro_embedded_bc}, we prescribe the
piezometric head on the Dirichlet boundary on the top and Neumann no-flow on the remaining sides of the rectangular box. 
The permeability is set equal to $\mathbf{K}_{\gamma}=10^{-6}\,\text{[m/s]}$ for the fractures and $\mathbf{K}=10^{-8}\,\text{[m/s]}$ for the matrix. Fracture aperture is about $\epsilon_1=10.16$ for the left fracture and about $\epsilon_2=15$ for the right fracture.


We use a gradient recovery strategy for the a-posteriori error estimation to guide the adaptive refinement (\text{AR}) of the matrix and the fracture meshes, and compare the numerical results with those computed employing uniform refinement (\text{UR}).  To this aim, we estimate the accuracy of each numerical simulation by computing the error with respect to the reference solution obtained by~\citet{flemisch2018} using a mimetic finite difference (MFD) method on a very fine mesh with $2\,260\,352$ triangles and $3\,471\,040$ dofs.  

In Tables~\ref{tab:tableHydroEmbeddedU} and~\ref{tab:tableHydroEmbeddedA} we report the results related to the embedded-ED technique for $\text{UR}$ and \text{AR} test cases, respectively. 
In particular, we specify the number of elements in the matrix mesh ({\bfseries \#-matr.}) and in the fracture mesh ({\bfseries\#-frac.}), the number of degrees of freedoms ({\bfseries{d.o.f}}), the density of non zeros entries ( {\bfseries{nnz/size$^2$}}), the condition number ({\bfseries $\| \cdot \|_2$-cond}), and the error computed with respect to the reference solution ($\textbf{err}_m$). In Table~\ref{tab:tableHydroEmbeddedU} we also specify the number of uniform refinements ($\#\textbf{UR}$), whereas 
in Table~\ref{tab:tableHydroEmbeddedA} we report the number of adaptive refinements ($\#\textbf{AR}$), and the error-threshold used for the gradient recovery strategy ($\textbf{err}_g$).
We point out that both the $\text{UR}$ and the $\text{AR}$ strategies are only performed on the matrix mesh. 

In Fig.~\ref{fig:hydro_embedded_mesh_amr} we show the meshes obtained for all the six $\text{AR}$ test 
cases. The $\text{AR}$ based on the gradient recovery strategy is combined with an $\text{AR}$ perform  ed on the overlapping region between the matrix and the fracture network.  The number of $\text{AR}$ performed in the overlapping zone are reported in Table~\ref{tab:tableHydroEmbeddedA} and specified within parentheses.

From the results summerized in Tables~\ref{tab:tableHydroEmbeddedU} and~\ref{tab:tableHydroEmbeddedA},  we observe that the error $\textbf{err}_m$ progressively reduces by increasing the number of refinements. However, the adaptive strategy allows accurate results  with coarser meshes when compared to the  uniform technique. While a uniform refined mesh with more than $200\,000$ elements ({\bfseries \#-matr.}) is needed to get an error ($\textbf{err}_m$) close to $10^{-6}$, an adaptively refined mesh with a total number of elements ten times smaller ensures the same accuracy.

Fig.~\ref{fig:hydro_embedded_ED}
shows the piezometric head obtained over line \aaline{} for the UR (a) and the AR (b) test cases. One may note large discrepancies in the area of the left fracture, especially for the results referred to the uniform refined meshes. In this regard, we found that the a-posteriori error  $\textbf{err}_g$ was higher close the sharp top boundary and in the zone occupied by the left fracture. Indeed, Fig.~\ref{fig:hydro_embedded_mesh_amr} (3)-(5) show that the $\text{AR}$ strategy produces a mesh size reduction in such regions. Hence, all the $\text{AR}$ test cases reveal a better agreement with the reference solution when compared to the $\text{UR}$ scenarios.

For completeness, the numerical results obtained for the test case $3\,\text{AR}$  are presented in Fig.~\ref{fig:hydro_embedded_bc}. Here, the colour refers to the spatial distribution of the piezometric head, whereas the contour lines are in black.

We perform the same analysis for the mortar-ED scenario and collect all the numerical results in Tables~\ref{tab:tableHydroEmbeddedU} and~\ref{tab:tableHydroEmbeddedA}. 
Again, the use of $\text{AR}$ allows us to achieve an accuracy comparable to the $\text{UR}$ test cases with less elements  in both the matrix mesh ({\bfseries \#-matr.}), and  the fracture mesh  ({\bfseries \#-frac.}).

In Fig.~\ref{fig:hydro_embedded_MR} we observe that the piezometric head profile computed along line \aaline{} converges to the reference solution (MFD) by increasing the number of refinements for all the numerical simulations. 
We point out that, while the embedded-ED test cases reveal large discrepancies  with respect to the reference solution in the region of the left fracture, only little differences are observed  for the mortar-ED scenarios. Indeed, the mortar approach requires a matching geometry at the interface between  fracture and  matrix, and consequently more accurate results are achieved even with coarser meshes.


\subsubsection{2D Benchmark Case 3: Realistic fracture network}\label{sec:real}
We consider the same settings and reference solution used in~\citet{flemisch2018} \emph{Benchmark 4}.
For this benchmark, a more realistic fracture network is taken into consideration. Here, the size of the domain is 700\,m x 600\,m with a fracture network of 64 fractures divided in 13 connected groups. The matrix permeability is $\mathbf{K} = \mathbf{I}  \cdot 10^{-14}\;$m$^2$
All the fractures have the same permeability $10^{-8}\;$m$^2$ and the same aperture $\epsilon = 10^{-2}\;$m. Hence, we have $\mathbf{K}_{\gamma^1} = \epsilon \mathbf{I}  \cdot 10^{-8}$ for the HD case, and $\mathbf{K}_{\gamma^2} = \mathbf{I}  \cdot 10^{-8}$ for the ED one.
There are no-flow boundary conditions on top and bottom of the domain. 
A pressure of $1\,013\,250$ Pa is imposed on the left boundary and of 0 Pa on the right boundary. 
Due to the rather complex geometry of the fracture network, the mortar method is impractical, hence it is not used here. We exclusively present results for the equi-dimensional embedded method. The mesh of the matrix has been automatically refined at the intersection with the fracture, for each experiment with 3, 6, and 7 levels of non-conforming mesh refinement, respectively. It can be observed in Table~\ref{tab:real} the condition numbers have comparable magnitudes for both HD and ED versions.
The condition number becomes large with the extremely varying mesh size resulting from the heavily focused refinement around the fractures. 

\begin{figure}[ht]\centering\footnotesize
    \parbox{\linewidth} {
    \includegraphics[width=0.45\linewidth]{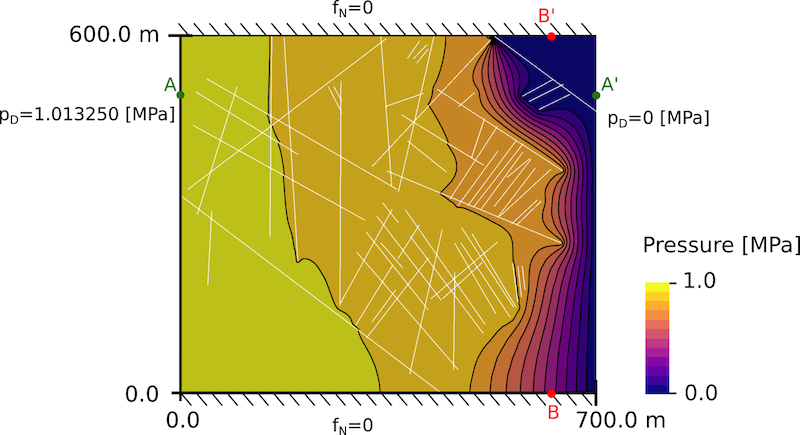} \hfill
    \includegraphics[width=0.25\linewidth]{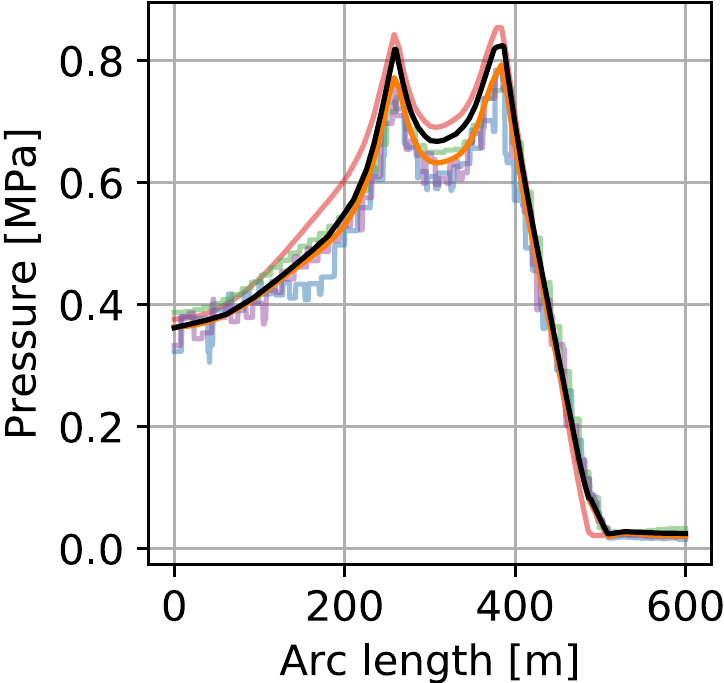} \hfill
    \includegraphics[width=0.25\linewidth]{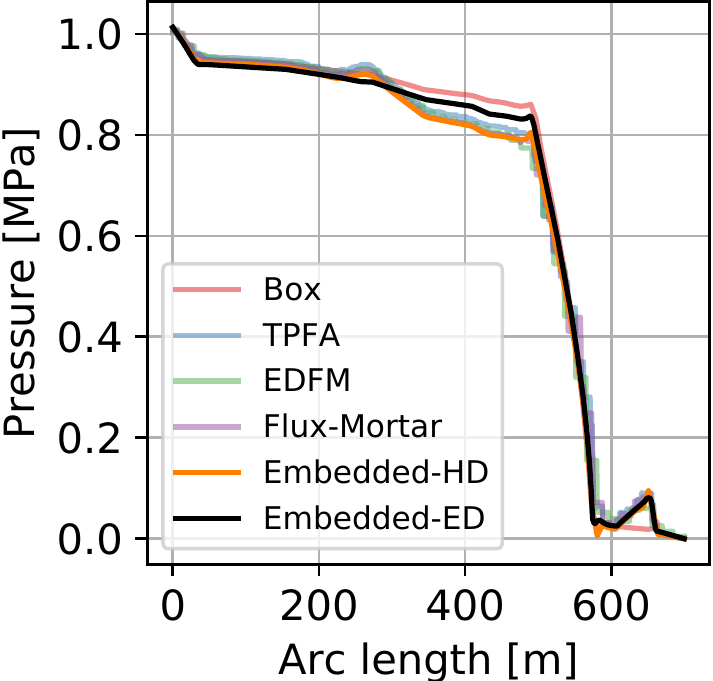}
    } \\[2em]
    \parbox{\linewidth}{
        \parbox{0.45\linewidth}{\centering (a) Pressure solution 
        } \hfill
        \parbox{0.25\linewidth}{\centering (b) Line \aaline{} } \hfill
        \parbox{0.25\linewidth}{\centering (c) Line \bbline{} }
    }
    \caption{
    \emph{Benchmark 4} in~\citet{flemisch2018}. Pressure solution for real fracture network with 64 fractures (a). Pressure profiles along the lines (b) \aaline and (c) \bbline. for both Embedded-HD and Embedded-ED  techniques.
    }
    \label{fig:realembedded}
\end{figure}

As can be observed in Fig.~\ref{fig:realembedded}, the results obtained are similar to those obtained by other methods in the field although no reference solution is available. 
It can be noticed that the equi-dimensional variant is closer to the \emph{Box} method compared to most other methods, including the hybrid dimensional results presented in~\cite{SCHADLE201942}.
From a practical perspective the equi-dimensional technique is slightly more complex since the fractures are extruded in normal direction, although automatically, thus requiring many more elements.

\todo{
\begin{table}
    \caption{%
    \emph{Benchmark 4} in~\citet{flemisch2018}. Mesh information, characteristics of system matrix and error in the matrix for the \emph{2D Realistic fracture network} benchmark in Embedded-ED and Embedded-HD cases.  
 }
    \input{\tab results_2DReal.tex}
    \label{tab:real}
\end{table}
}

%
\subsection{3D Experiments}
Flow through fractured porous media is largely governed by 3D effects. 
Therefore, this section presents an application and evaluation of the dual Lagrange multiplier methods in 3D.
First, results obtained with three Lagrange multiplier methods (embedded HD, embedded ED, mortar HD) are compared to results of~\num{17} methods presented in the benchmark study \citet{berre_2020}.
More complex benchmark cases in 3D are studied for the embedded HD method as part of the aforementioned benchmark study. 
The embedded HD method is preferably used for those geometrically complex cases as it eases meshing of the fracture networks and the porous matrix mesh can be chosen regular. 
Fracture network mesh generation for the embedded ED method is very challenging, which is particularly relevant at the fracture intersection.
Furthermore, all 3D benchmark cases in~\citet{berre_2020} are designed for very small fracture apertures and therefore equi-dimensional fracture meshes result in poor mesh quality. 
It is also important to bear in mind that equi-dimensional fracture meshes are mainly necessary for large aperture values and the benchmark cases yield no reason to use equi-dimensional meshes. 
Furthermore, the mortar method is less suited for complex fracture geometries due to the complexity of both, setting up the mesh and dealing with over-constrained scenarios for the mortar conditions.
In a final realistic scenario the strength of each method is demonstrated and they are applied in a combined scenario as it could be typical for fractured systems.
\subsubsection{3D Benchmark Case 1: Single fracture}
\begin{figure}[hptb]
    \centering\footnotesize
    \includegraphics[width=.5\linewidth]{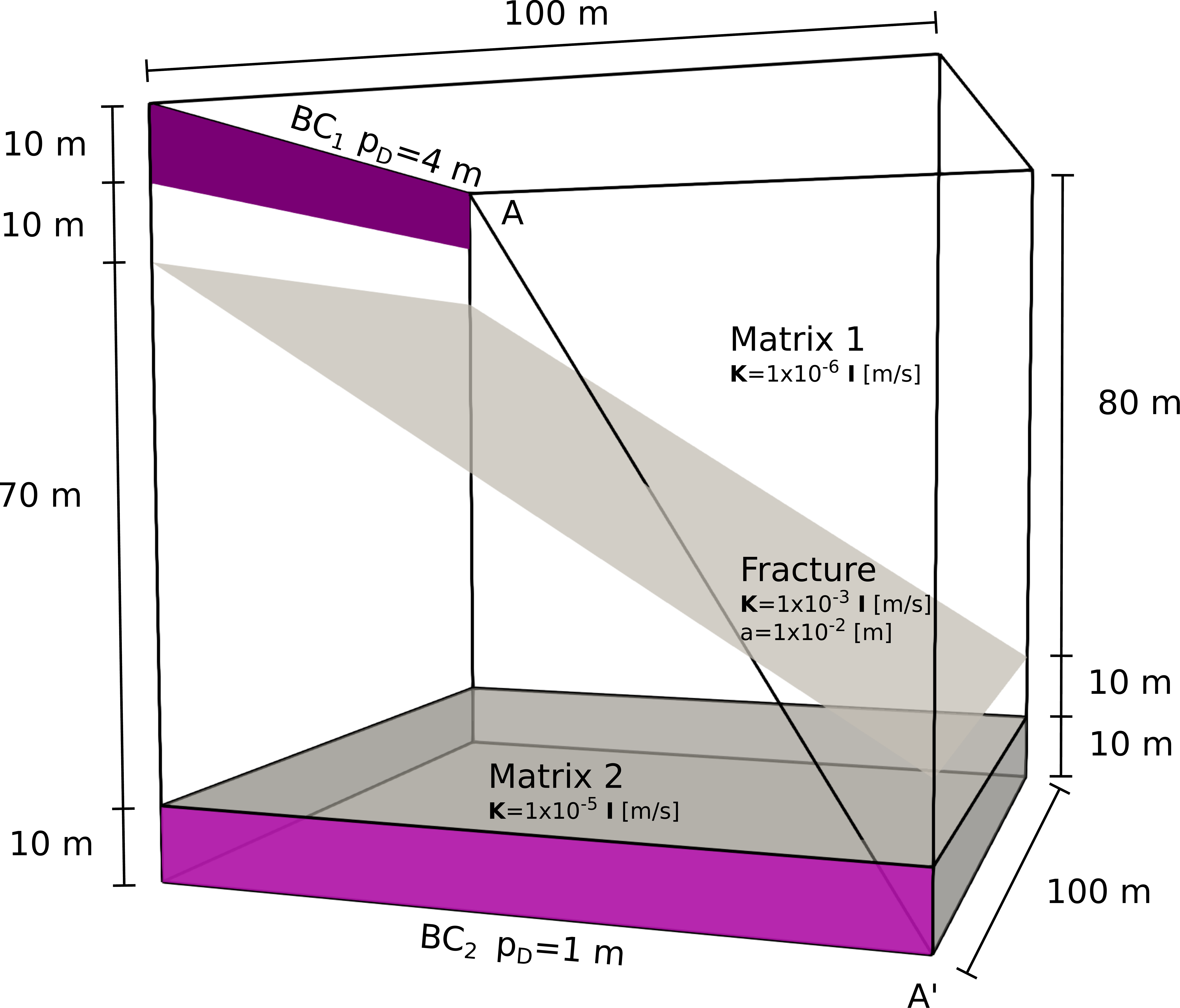}
    \caption{%
    Model domain (outlines) of benchmark \emph{Case 1: Single Fracture} in~\citet{berre_2020} with Matrix~1 and~2 and a single fracture intersecting Matrix~1.
    Boundary~1 (BC$_\text{1}$) is the inflow and boundary~2 (BC$_\text{2}$) the outflow.
    Results are compared along the line~\aaline{} through the matrix domains.
    }
    \label{fig:3D_BC1_domain}
\end{figure}

\begin{figure}[hptb]
    \centering\footnotesize
    \includegraphics[width=1.\linewidth]{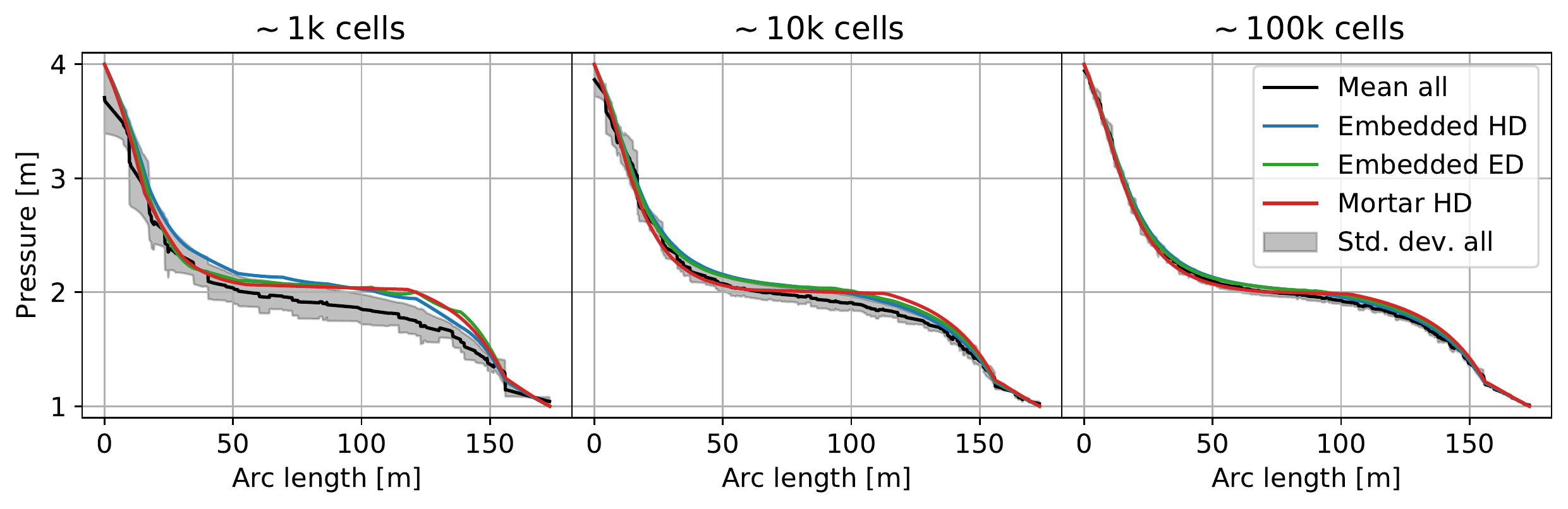}
    \includegraphics[width=1.\linewidth]{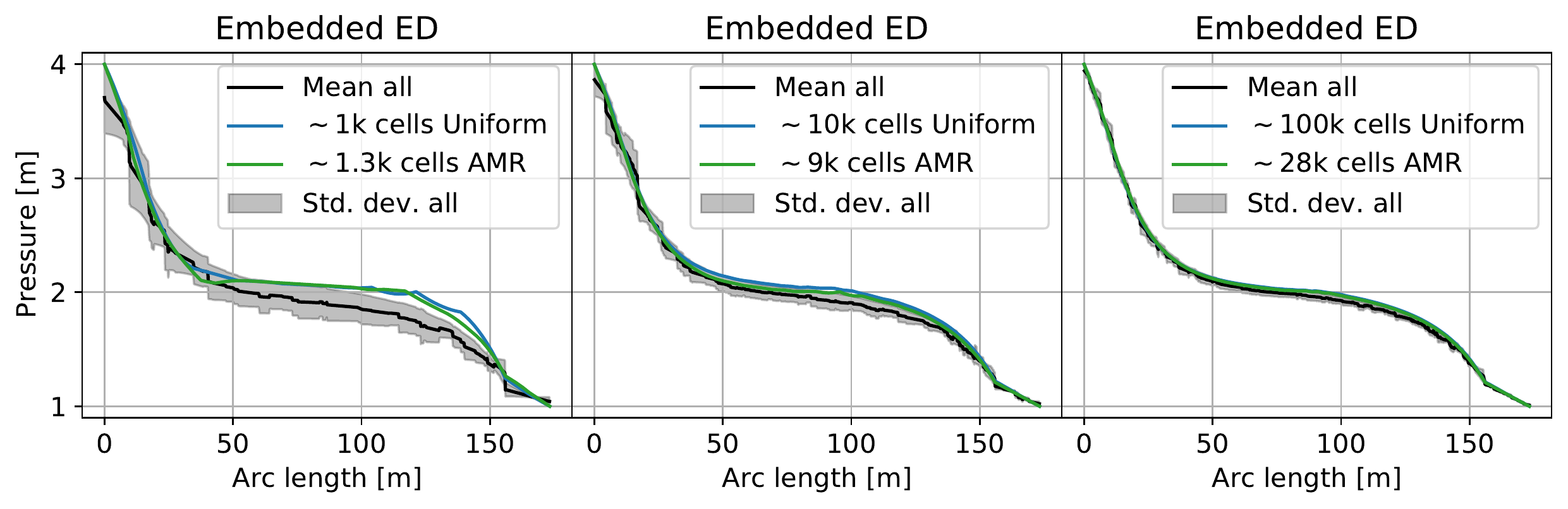}
    \caption{%
    Pressure results along the line~\aaline{} for $\sim$1k, $\sim$10k, and $\sim$100k matrix cells for uniform refinement (upper figure) and $\sim$1.3k, $\sim$9k, and $\sim$28k for adaptive mesh refinement (lower figure).
    Results are shown with uniform refinement for the hybrid-dimensional (HD) and equi-dimensional (ED) embedded method and the hybrid mortar method and with adaptive mesh refinement for the equi-dimensional embedded method.
    The black line shows the mean results of all benchmark methods and the grey range shows the standard deviation, respectively~\cite{berre_2020}. 
    }
    \label{fig:3D_BC1_results}
\end{figure}

In this section three Lagrange multiplier methods are compared to benchmark \emph{Case 1: Single Fracture} presented by~\citet{berre_2020}, which already includes results obtained with the hybrid embedded method, presented in~\citet{SCHADLE201942}.
Nevertheless, for completeness and for better comparison of the methods presented in this study the results of the hybrid embedded approach are again explicitly presented. 
To ease comparison, the results presented in the benchmark study are summarized and only the mean of all results as well as the standard deviation are plotted. 
The comparison of vastly different discretization methods is enabled by interpolating each solution to~\num{1000} evenly spaced points along a line. 
Following this, the mean and standard deviation are computed at each of these points.
It is important to note that the mean of all results is not necessarily the correct solution and just provides a measure of comparison to the results presented in the aforementioned benchmark study.
For the detailed benchmark results the interested reader is referred to \citet{berre_2020}.
Finally, improvements of the accuracy can be demonstrated for the embedded ED approach by adaptive mesh refinement.\\
Figure~\ref{fig:3D_BC1_domain} shows the model domain of benchmark \emph{Case 1: Single Fracture} which is adapted from \citet{zielke_1991} and \citet{barlag_1998}.
This case consists of a single fracture intersecting a matrix cube with~\SI{100}{\metre} edge length.
The lowest~\SI{10}{\metre} thick layer of the matrix block (Matrix~\num{2}) has an increased permeability. 
Fluid injection occurs at a Dirichlet boundary condition (BC$_\text{1}$) located above the fracture at the upper most~\SI{10}{\metre} thick layer of the matrix block.
A Dirichlet boundary condition (BC$_\text{2}$) which acts as the outflow, is located at Matrix~\num{2}.
The injection pressure is fixed at~\SI{4}{\metre} and the production pressure at~\SI{1}{\metre}.
The permeability~$\mathbf{K}$ in the Matrix~\num{1} is~$\mathbf{I}\cdot 10^{-6}$~[m/s] and in the Matrix~\num{2}~$\mathbf{I}\cdot 10^{-5}$~[m/s].
The fracture has a permeability $\mathbf{K}_{\gamma}$ of~$\mathbf{I}_{\gamma}\cdot 10^{-3}$~[m/s] with an aperture $\epsilon$ of~$10^{-2}$~[m].
%
\\
%
The results obtained with the hybrid- and equi-dimensional embedded method and the hybrid mortar method are presented in Figure~\ref{fig:3D_BC1_results}. 
In the upper part of the figure, three different mesh sizes ($\sim$1k, $\sim$10k, and $\sim$100k matrix cells) are compared to~\num{17} methods presented in~\citet{berre_2020}.
It is important to keep in mind that the hybrid-dimensional embedded results are also part of the benchmark results.
The pressure solution for all methods is compared along the line~\aaline{} (see Fig.~\ref{fig:3D_BC1_domain}). \\
Overall, the results of the methods presented here and the methods presented in~\citet{berre_2020} show good convergence towards a common solution.
Particularly for the very coarse mesh of only~$\sim$1k cells in the matrix domains the Lagrange multiplier methods show some deviations. 
These deviations are more pronounced for the embedded HD method at the first half of line~\aaline{} and for the embedded ED and the mortar method at the second half. 
The deviations are partially due to the fact that most of the other methods presented in the benchmark study represent the fracture geometry in the matrix explicitly in matching mesh geometries.
Embedded techniques typically require a resolution that is roughly twice as high as fitted mesh techniques to achieve the same accuracy.
Already for the case with~$\sim$10k cells the embedded HD and ED methods show very similar results. 
For the coarsest case neither of the methods is preferable as they all show deviations in different regions. 
Furthermore, the solution of the mortar method shows a kink at~$\sim$~\SI{115}{\metre} for all mesh sizes. 
As mentioned above, this is due to over-constrained dofs, which poses significant technical challenges to be automated in 3D. \\
As shown in~\ref{sec:hydrocoin} adaptive mesh refinement allows to obtain more accurate results while reducing the number of matrix elements.
Here the matrix mesh for the embedded ED approach is adaptively refined with up to two refinement steps, starting at~$\sim$500 cells by progressively reducing the threshold of error adopted for the gradient recovery strategy.
For brevity, we refer to the error threshold as $\mathbf{err}_g$ and to the number of  adaptive refinement steps as $\text{AR}$. 
Thus, we employ $\mathbf{err}_g=\num{15}$ and $\text{AR}=\rm{1}$ for the coarsest mesh, $\mathbf{err}_g=\num{5}$ and $\text{AR}=\num{2}$ for the middle mesh, and $\mathbf{err}_g=\num{0.1}$ and $\text{AR}=\num{2}$ for the finest test case.
One may note that the the error threshold, $\mathbf{err}_g$, is progressively  reduced to increase the accuracy of the numerical results. The matrix meshes resulting from the adaptive mesh refinement have~$\sim$1.3k, $\sim$9k, and $\sim$28k cells.
In Fig.~\ref{fig:3D_BC1_results}, second row, we compare the results obtained with uniformly refined meshes with those obtained with adaptive mesh refinement.
While the $\mathrm{amr}_1$ has some more cells than the compared uniform mesh, the pressure solution along parts of the line~\aaline{} is closer to the mean of all benchmark methods. 
For the two finer meshes, an improvement in accuracy is clearly visible.
Taken together, all Lagrange multiplier methods presented here show good convergence and match well with other methods presented in the benchmark study.
%
\subsection{3D Realistic scenario}
\begin{figure}[hptb]
    \centering\footnotesize
    \includegraphics[width=.5\linewidth]{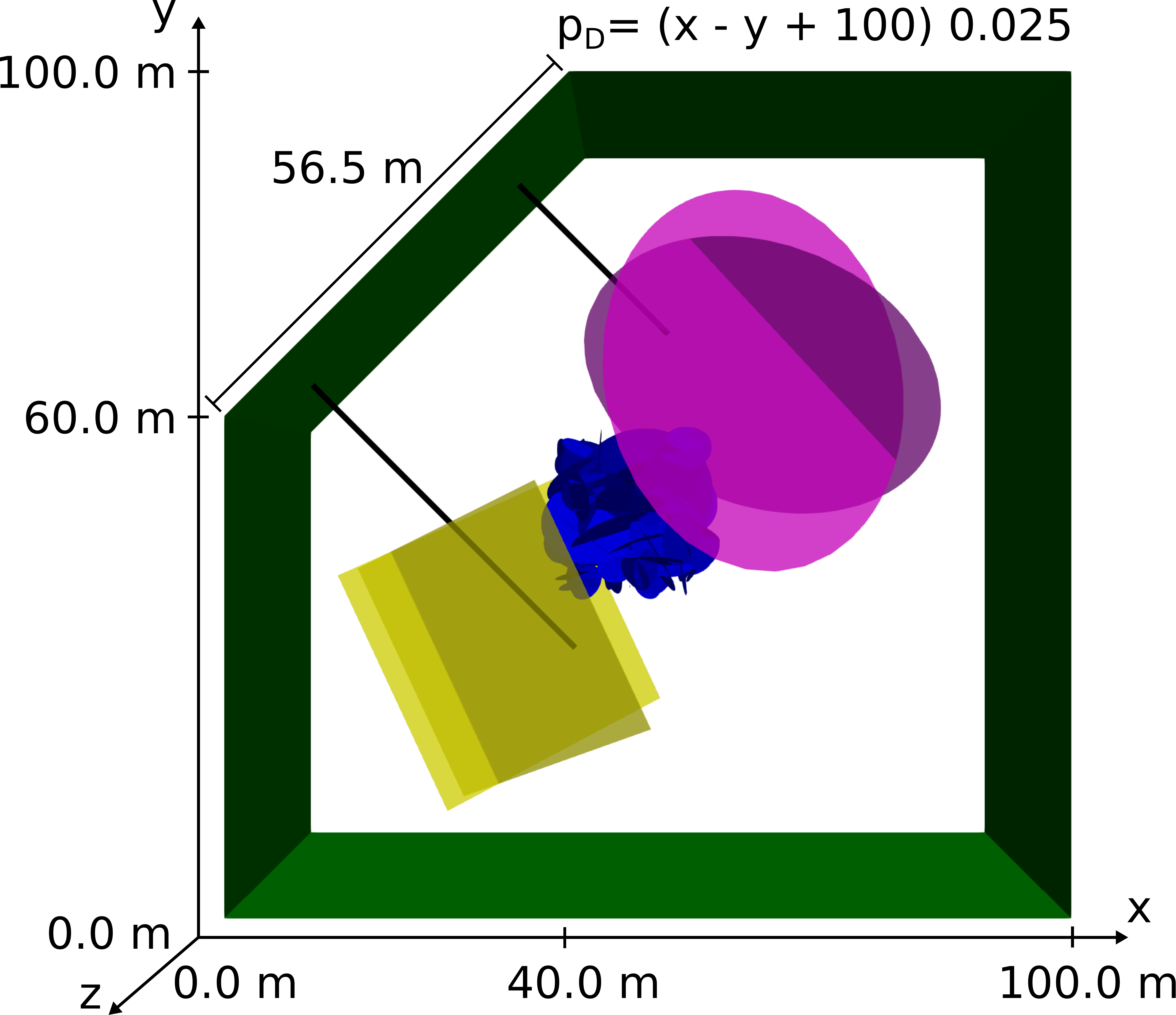}
    \caption{%
    Model domain of the complex case with a combined application of all presented methods.
    The upper right corner represents a ventilated access tunnel, as it is typical in deep underground laboratories. 
    Fractures computed with the hybrid embedded method are depicted in blue and fractures computed with the equi-dimensional embedded method in pink.
    The yellow domains, with low permeability, are equi-dimensional and coupled by the mortar method.
    The green planes show the outer boundaries and the black lines indicate the injection and production boreholes.
    }
    \label{fig:complex_case_domain}
\end{figure}

\begin{figure}[hptb]\centering\footnotesize
    \parbox{\linewidth} {
    \includegraphics[width=0.29\linewidth]{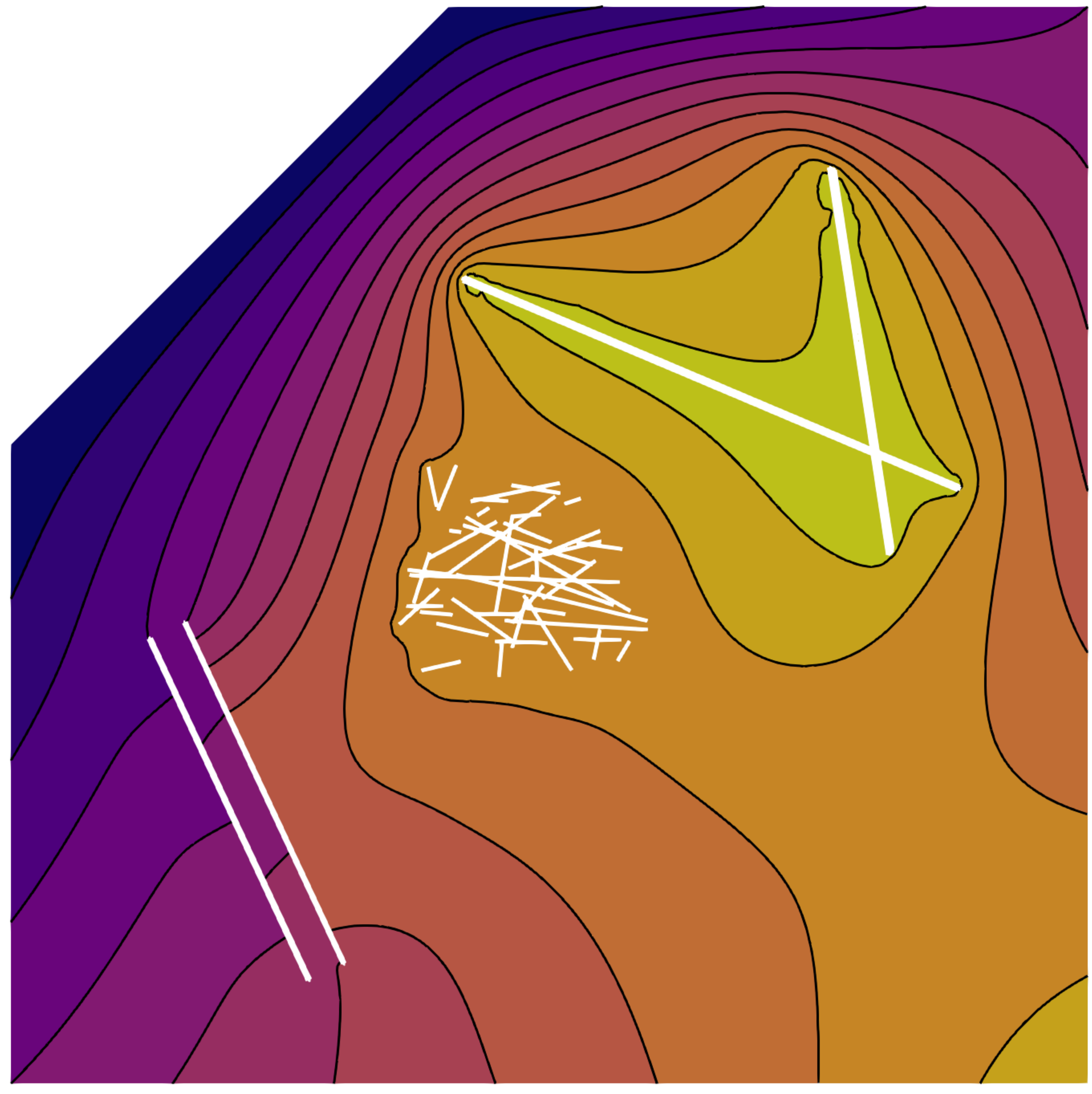} \hfill
    \includegraphics[width=0.29\linewidth]{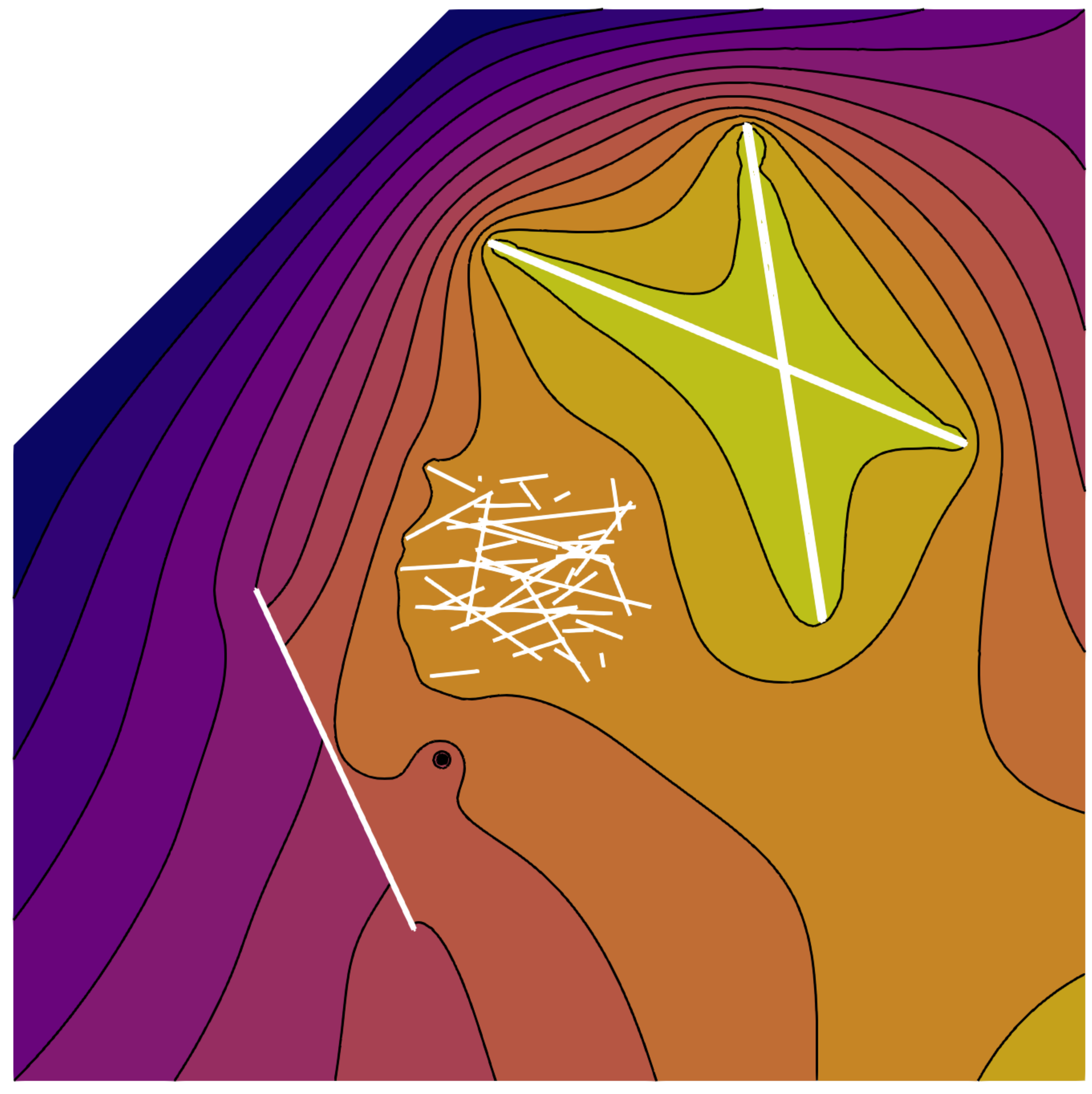} \hfill
    \includegraphics[width=0.29\linewidth]{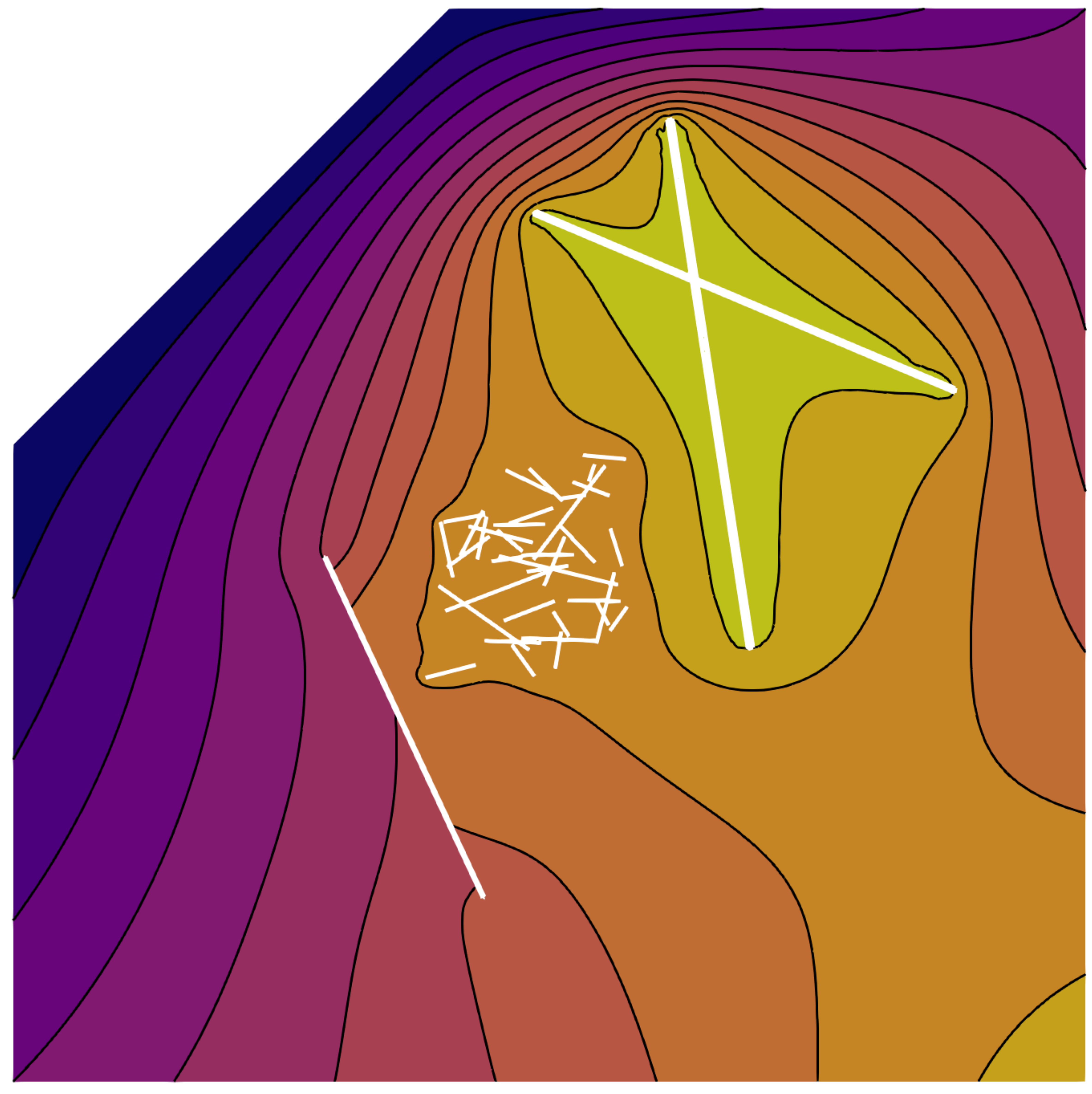} \hfill
    \includegraphics[width=0.09\linewidth]{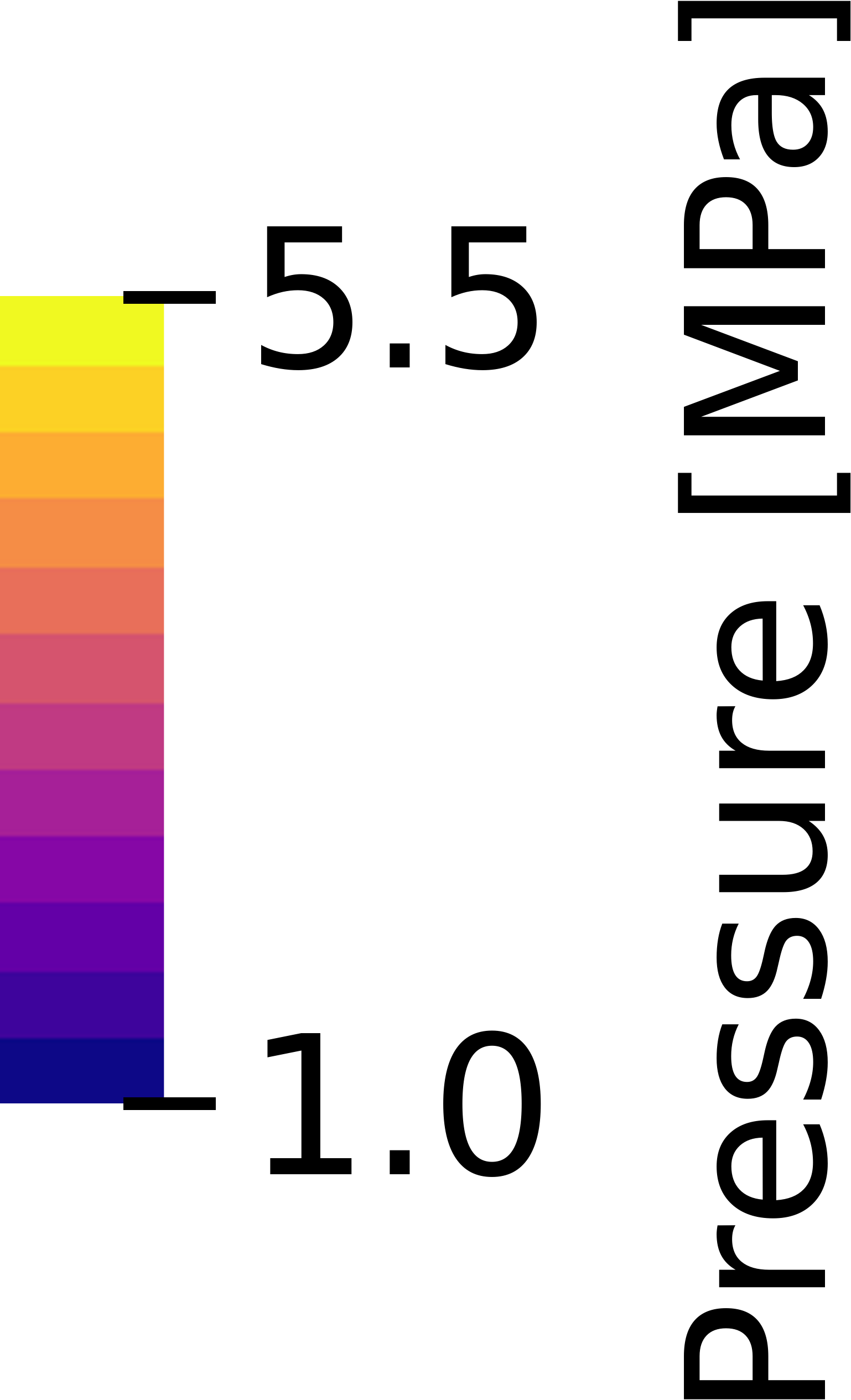} 
    } \\[1em]
    \parbox{0.87\linewidth}{
        \parbox{0.29\linewidth}{\centering (a) Plane at $z=$~\SI{-5}{\metre}} \hfill
        \parbox{0.29\linewidth}{\centering (b) Plane at $z=$~\SI{0}{\metre}} \hfill
        \parbox{0.29\linewidth}{\centering (c) Plane at $z=$~\SI{5}{\metre}} 
    }\hspace{0.09\linewidth}
    \caption{%
    Pressure solution of realistic scenario case at three planes intersecting the domain at~\SI{-5}{\metre},~\SI{0}{\metre}, and~\SI{5}{\metre} in \emph{z}-direction.
    }
    \label{fig:complex_case_results}
\end{figure}

Geological setting with fractures and faults often require to represent such features with permeability and aperture ranging over several orders of magnitude. 
Therefore, their combined representation in numerical models is crucial for a complete description of geological settings. 
The setup of present realistic scenario is loosely based on the geological setting at the Grimsel Test Site, (GTS)~\cite{amann_2018}.
However, it is important to note that the steady-state flow field, as computed here, is difficult to achieve in experiments conducted in such laboratories with a very low permeability rock matrix.
Nevertheless, the given setup allows to demonstrate the different strength of the particular methods described in this study, i.e. complex fracture networks, fractures with large aperture widths, and blocking fractures.
Even more importantly, this realistic scenario shows the integration of several coupling strategies in a single joint framework method. \\
Figure~\ref{fig:complex_case_domain} shows the model domain with a complex fracture network (blue) located between two large features with two intersecting fractures each.
One of these features acts as blocking fractures (yellow) with low permeability values, the fractures in the other feature have large apertures (magenta). 
Furthermore, two boreholes are drilled into the rock domain with one of them ending in the large aperture fractures and the other one in the rock domain.
The upper right corner of the square domain is not modeled as it acts as an access tunnel with atmospheric pressure.
Consequently, the pressure at the outer boundary is fixed by a Dirichlet boundary condition of $p_D=(x-y+100)\;0.025$~(green).
This boundary condition results in a pressure of~\SI{1}{\mega\pascal} at the access tunnel location and~\SI{5}{\mega\pascal} at the lower right corner. 
At the top and bottom (\emph{z}-direction) of the domain no-flow boundary conditions are applied.\\
In the numerical model the fractures and fracture network are represented by three different methods described above. 
The complex fracture network (blue) is meshed with lower-dimensional manifolds and embedded in the matrix mesh and has been initially presented by~\citet{SCHADLE201942}.
In this study the fracture aperture and permeability and the matrix permeability are chosen so that the fractures and the matrix contribute similarly to the overall flow.
In contrast, the rock matrix in the present experiment holds a low permeability and the fracture properties are chosen to dominate the flow.
The fracture radius distribution in the network follows a power law, with truncations at~\SI{2.5}{\metre} and~\SI{10}{\metre}. 
The~\num{150} fractures are circular, randomly oriented, and distributed in a cubical area of the model domain with a side length of~\SI{25}{\metre}.
The fracture aperture is~$\sim 10^{-4}$~m and the permeability chosen to be~$\sim 10^{-9}$~m$^{\text{2}}$.
Furthermore, the two fractures with large apertures (magenta) are represented by equi-dimensional domains embedded in the matrix domain. 
One of these fractures is intersected by a borehole which is explicitly meshed as a sub-domain.
The two fractures are circular with an aperture width of~\SI{3e-1}{\metre} and radii of~\SI{25}{\metre}.
Further, the infilling material of these fractures is assumed to have a permeability of~\SI{1e-12}{\metre\squared}.
In the borehole domain a forcing function of~\num{1.0e-13} is applied in a volume of~\SI{1e-2}{\cubic\metre}, acting as an injection borehole with a pressure of~$\sim$\SI{5.5}{\mega\pascal}.
To represent fractures with low permeability, acting as blocking fractures, two fractures (yellow) are described by equi-dimensional domains coupled to the matrix mesh by the mortar method. 
These fractures are rectangular with a side length of~\SI{35}{\metre} and permeability of~\SI{1e-21}{\metre\squared}.
Generally, the shape of the fractures might of any shape, e.g.\, circular or rectangular. 
The second borehole acts as a sink with a fixed atmospheric pressure. This borehole is described by a mortar coupling of a small domain with the dimensions of the borehole and a fixed pressure of~\SI{1}{\mega\pascal}.
Finally, the matrix permeability is~\SI{1e-18}{\metre\squared}.
\\
%
%
Figure~\ref{fig:complex_case_results} shows the pressure distribution across three planes intersecting the model domain parallel to the \emph{xy}-plane. 
Throughout all planes the low permeability fractures act as discontinuities for pressure. 
Moreover, for plane~(a) a clear discontinuity can be observed across the two fracture cross sections. 
With the high aperture and permeability of the complex fracture network and the low permeability of the rock matrix the pressure across the fracture network is equilibrated, connecting the upper right with the lower left part of the domain. 
With the injection borehole in one of the two equi-dimensional fractures, these fractures and the surrounding rock matrix area subject to the largest pressure values.
In plane~(b) the production borehole locally reduces the pressure to atmospheric pressure. 
However, due to the low permeability of the rock matrix the gradient around this borehole is very steep and the influence on the overall solution is limited.
Additionally, the fixed pressure boundary condition at the outer boundary forces the pressure to steep gradients close to the embedded ED fractures with high injection pressure. 
These steep gradients result from boundary effects and for a representative study of such a geological setting the domain would have to be extended in these areas. 
However, the goal of this study is to demonstrate the application of Lagrange multipliers for different coupling strategies, spanning from equi-dimensional to lower dimensional models, and from embedded to mortar techniques.
Ultimately allowing for more flexibility in the treatment of the fracture and matrix configuration.
Furthermore, such steep gradients are generally difficult to resolve  in numerical models, thus adding further complexity to this test case.
In summary, this realistic scenario demonstrates the strength of each method and the ability of the presented unified framework to combine all of these approaches while yielding smooth pressure results in geologically complex settings. 
More specifically, the hybrid-dimensional embedded approach eases meshing of complex fracture networks of fractures with small aperture widths, the equi-dimensional embedded approach allows to consider fractures with large aperture widths, and the equi-dimensional mortar approach enables to consider blocking fractures.
\\
%
\section{Conclusion} \label{sec:conclusion}

This study expands on previous works based on the application of the Lagrange multiplier method to compute single-phase fluid flow problems in fracture dominated porous media. 
In particular, we employ the finite element method in combination with an $L^2$-projection operator to couple different types of non-conforming meshes. 
The non-conformity might arise at the interface of independently meshed sub-domains, at hanging nodes resulting from adaptive mesh refinement, or by combining multiple overlapping meshes. 
Furthermore, fractures are either described by equi-dimensional or hybrid-dimensional domains. 
The applied Lagrange multiplier is discretized using dual basis functions, which provide two main advantages.
First, the number of degrees of freedom is reduced to the ones of the background mesh representing the porous-matrix.
Second, the arising symmetric-positive-definite linear systems are convenient work with. 
\\
Here, we present a unified framework covering all coupling techniques mentioned above.
The different mesh solutions are compared with state-of-the-art benchmark cases in 2D and 3D, the numerical performance is studied, and use cases are presented in isolation and combination.
Overall, the results suggest that the presented tool-set is capable of computing fluid-flow through complex and heterogeneous rock formations in a robust and convenient way.
It is important to bear in mind that realistic scenarios of fracture dominated rock formations may include fractures with geometric and physical properties ranging over many orders of magnitude.
Therefore, the presented single joint framework allows to deeply exploit non-conforming hybrid- and equi-dimensional fracture models and efficiently combine these models. 
By using the dual Lagrange multiplier we are able to combine multiple complex fracture networks without changing the size of the algebraic system arising from the porous-medium matrix, although we have more non-zero entries associated with the coupled degrees of freedom. It is also worth to point out that the conditioning of the system is not worsened, as it is the case when using other types of Lagrange multipliers~\cite{SCHADLE201942}.  
\\
With the integration of adaptive mesh refinement in the solution process, the error in the solution can be controlled in an automated way either by means of an error estimator or by pre-defining areas of interest for refinement. This allowed to complement the discussion in~\citet{SCHADLE201942} about the necessity of having a finer mesh around fractures and their tips and intersections. 
\\
The study of a realistic geological setting, inspired by the Grimsel Test Site, demonstrates the advantage of the unified framework to represent vastly different fracture geometries and properties within a single numerical model. 
This allows to expand on existing numerical studies of such systems by the opportunity to include a large range of fracture representations.
Furthermore, the highly efficient and convenient tools combined with the eased meshing of non-conforming meshes enables stochastic studies for a wide range of fracture dominated systems.
\\
Further investigations based on this unified framework would benefit by focusing on mass conservation properties of the finite element discretization and application to transport problems.

%
\section*{Acknowledgment}
P.Z., M.G.C.N., and L.K. thank the SCCER-SoE and SCCER-FURIES programs, and the PASC project FASTER: Forecasting and Assessing Seismicity and Thermal Evolution in geothermal Reservoirs.
P.S. thanks the Werner Siemens Foundation for their endowment
of the Geothermal Energy and Geofluids group at the Institute of Geophysics, ETH Zurich. 

\section*{Authorship statement}
    P.Z. lead the drafting of the manuscript, implemented most methods and numerical tools, developed parts of the conceptual models, and lead parts of the numerical experiments.
    P.S. contributed drafting the manuscript, lead the development of the 3D conceptual models, their validation, and presentation of the results.
    L.K. produced, collected, and prepared most of the 2D benchmark results.
    M.G.C.N contributed drafting the manuscript, implemented the adaptive refinement strategy and its integration with the variational transfer, lead the numerical experiments related to the  adaptive refinement, and contributed to the 2D and 3D numerical experiments.

\section*{Conflict of interest}
The authors declare that they have no conflict of interest.

%
\section*{Computer Code Availability}
All methods and routines, used for this study, are implemented with the open-source software library \emph{Utopia}~\cite{utopiagit}. \emph{Utopia}'s lead developer is author Patrick~Zulian at USI~Lugano, Switzerland. 
The contact address and e-mail of Patrick Zulian are as follows:
\begin{itemize}
\item[] Institute of Computational Science\\
Universit{\`a} della Svizzera italiana (USI - University of Lugano)\\
Via Giuseppe Buffi 13\\
CH-6900 Lugano
\item[] patrick.zulian@usi.ch
\end{itemize}
\emph{Utopia} was first available in 2016, the programming language is \emph{C++} and it can be accessed through a git repository or a docker container on:
\begin{itemize}
\item[] \url{https://bitbucket.org/zulianp/utopia} (approx. 40~MB of uncompressed data),
\item[] \url{https://hub.docker.com/r/utopiadev/utopia}.
\end{itemize}
The software dependencies are as follows:
\begin{itemize}
\item[] PETSc (\url{https://www.mcs.anl.gov/petsc}),\\ must be compiled with \emph{MUMPS} enabled
\item[] libMesh for the FE module (\url{https://github.com/libMesh})
\end{itemize}
There are no hardware requirements given by \emph{Utopia}.
Potential hardware or software requirements of the underlying libraries \emph{libMesh} and \emph{PETSc} are not stated here. 

\bibliographystyle{unsrtnat}
\bibliography{references}

\end{document}

%% file: tables/results_2DRegular.tex
\scriptsize
\centering
\begin{tabular}{llrrllll}
\toprule\bfseries Method & \bfseries \#-matr. & \bfseries \#-frac. & \bfseries d.o.f. & \bfseries nnz/size$^2$ & \bfseries $\| \cdot \|_2$-cond. & \bfseries err$_m$ \\\midrule
Embedded-ED & 10 656 triangles & 8648 & 5853 & 1.3e-3 & 3.7e6 & 3.3e-5\\
Embedded-ED & 47 142 triangles& 8648 & 25\,992 & 2.8e-4 & 1.8e8 & 1.9e-6\\
Embedded-ED & 96 762 triangles& 8648 & 53\,379 & 1.4e-4 &  1.2e9 & 2.8e-7\\
\hline
Mortar-ED & 12 791 triangles& 34 592 & 32\,935 &  2.3e-4 & 5.2e10 & 6.4e-8\\
Mortar-ED & 36 331 triangles& 34 592 & 45\,172 & 1.7e-4 & 8.8e10 & 2.8e-8\\
\hline
Mortar-ED (blocking) & 922 triangles& 34 592 & 26\,520 & 2.7e-4 & 9.4e7 & 3.3e-8\\
\hline
Embedded-HD & 1089 quads& 112 & 1156 & 9.7e-3 & 3.0e4 & 9.7e-3\\
Embedded-HD & 16 641 quads & 448 & 16\,900 & 5.6e-4 & 1.6e6 & 2.5e-3\\
Embedded-HD & 66 049 quads & 896 & 66\,564 &  1.4e-4 & 1.3e7 & 1.3e-3\\

\bottomrule
\end{tabular}

%% file: tables/results_2DHydroCoinUR.tex
\scriptsize
\centering
\begin{tabular}{llrrllll}
\toprule\bfseries Method & \bfseries \#-matr. & \bfseries \#-frac. & \bfseries d.o.f. & \bfseries nnz/size$^2$ & \bfseries $\| \cdot \|_2$-cond. & \bfseries err$_m$ & \#\textbf{UR} \\\midrule
Embedded-ED & 272 triangles & 3 480 &  159 & 4.3e-2 & 2.9e9 & 5.1e-4& 1\\
Embedded-ED & 1 088 triangles&  3 480 & 589 & 1.2e-2 & 1.1e10 & 8.5e-5&2\\
Embedded-ED  & 4 352 triangles&  3 480 & 2 265 & 3.1e-2 &  4.3e10 & 1.7e-5&3\\
Embedded-ED & 17 408 triangles&  3 480 &  8 881 &  8.1e-4 & 1.8e11 & 4.0e-6& 4\\
Embedded-ED & 69 632 triangles&  3 480& 35 169 & 2.1e-4 & 6.8e11 & 1.9e-6 & 5\\
Embedded-ED & 278 528 triangles & 3 480 & 139 969 & 5.5e-5 & 2.7e12 & 1.3e-6 & 6\\
\hline
Mortar-ED & 1 116 triangles & 220 & 246 &  9.7e-3& 3.6e6 &2.1e-4 & 1\\
Mortar-ED & 4 464 triangles & 880 & 2 973 & 2.5e-3 & 7.6e6 & 1.9e-4 & 2 \\
Mortar-ED & 17 856 triangles & 3520 & 11\,285 &  6.4e-4 & 2.7e7 & 2.0e-4 & 3\\
Mortar-ED &  71 424 triangles & 14 080 & 43 941 &  1.7e-4 & 1.1e8 & 2.0e-4 & 4\\
\bottomrule
\end{tabular}

%% file: tables/results_2DHydroCoinAMR.tex
\scriptsize
\centering
\begin{tabular}{llrrllllll}
\toprule\bfseries Method & \bfseries \#-matr. & \bfseries \#-frac. & \bfseries d.o.f. & \bfseries nnz/size$^2$ & \bfseries $\| \cdot \|_2$-cond. & \bfseries err$_m$ &\bfseries \# AR & \bfseries err$_g$ \\\midrule
Embedded-ED & \,611 triangles & 3 480 &  352 & 2.0e-2 & 1.5e10 & 8.9e-5 & 0 (2) & -\\
Embedded-ED  & 1 544 triangles&  3 480 & 874 & 7.8e-3 & 5.3e10 & 1.6e-5 & 1 (2) & 5\\
Embedded-ED  & 2 378 triangles&  3 480 & 1354 & 4.5e-3 &  8.1e10 & 7.2e-5 & 2 (2) & 5\\
Embedded-ED  & 7 913 triangles&  3 480 &  4371 &  1.6e-3 & 3.5e11 & 2.9e-6 & 4 (2) & 1\\
Embedded-ED & 9 147 triangles&  3 480& 4991 & 1.5e-3 & 2.8e11 & 2.0e-6 & 4 (4) & 1\\
Embedded-ED & 21 599 triangles & 3 480 & 11658 & 1.02e-3 & 1.06e12 & 1.4e-6 & 4 (6) & 1\\
\hline
Mortar-ED & 2 127 triangles & 202 & 1375 & 5.9e-3 & 3.7e11 & 2.1e-4 & 1 (1) & 5\\
Mortar-ED & 8004 triangles & 332 & 2791 & 1.1e-3 & 1.9e12 & 1.9e-4 & 2 (2) & 5\\
Mortar-ED & 19 515 triangles & 442 & 10 600 & 7.1e-4 & 1.2e12 & 1.9e-4 & 3 (2) & 5\\
Mortar-ED&  28 692 triangles  & 457 & 15 284 &  4.8e-4 & 1.7e12 & 1.9e-4 & 4 (2) & 5\\
\bottomrule
\end{tabular}

%% file: tables/results_2DReal.tex
\scriptsize
\centering
\begin{tabular}{llrrllll}
\toprule\bfseries Method & \bfseries \#-matr. & \bfseries \#-frac. & \bfseries d.o.f. & \bfseries nnz/size$^2$ & \bfseries $\| \cdot \|_2$-cond.\\\midrule
Embedded-ED & 1405 quads & 199 996 & 1487 &  1.9e-4 & 2.0e15\\
Embedded-ED & 39 085 quads & 199 996 & 41\;773 &  2.3e-4 & 2.2e17\\
Embedded-ED & 94 513 quads & 199 996 & 102\;486 & 9.1e-5 & 5.8e17\\
\hline
Embedded-HD & 4200 quads & 1024 & 4331 & 3.8e-3 & 1.3e16\\
Embedded-HD & 67 200 quads & 4096 & 67\,721 & 1.6e-4 & 2.1e17\\

\bottomrule
\end{tabular}